\documentclass[fleqn,usenatbib]{mnras}

\usepackage{newtxtext,newtxmath}

\usepackage[T1]{fontenc}

\DeclareRobustCommand{\VAN}[3]{#2}
\let\VANthebibliography\thebibliography
\def\thebibliography{\DeclareRobustCommand{\VAN}[3]{##3}\VANthebibliography}

\usepackage{graphicx}	
\usepackage{amsmath}	
\usepackage{caption}
\usepackage{subcaption}
\usepackage{multirow}
\usepackage{float}

\RequirePackage[normalem]{ulem} 
\RequirePackage{color}\definecolor{RED}{rgb}{1,0,0}\definecolor{BLUE}{rgb}{0,0,1} 
\providecommand{\DIFdel}[1]{}

\title[
Joint astrophysical constraints on 21-cm models
]
{Constraining the properties of Population III galaxies with multi-wavelength observations}
\author[S. Pochinda et al.]
{S. Pochinda$^{1,2}$\thanks{E-mail: sp2053@cam.ac.uk},
T. Gessey-Jones$^{1,2}$,
H. T. J. Bevins$^{1,2}$,
 A. Fialkov$^{2,3}$,
 S. Heimersheim$^{2,3}$,
\newauthor I. Abril-Cabezas$^{2,4}$,
 E. de Lera Acedo$^{1,2}$,
S. Singh$^{5}$, 
S. Sikder$^{6}$, and 
R. Barkana$^{6}$
\\
$^{1}$Astrophysics Group, Cavendish Laboratory, J. J. Thomson Avenue, Cambridge, CB3 0HE, UK\\
$^{2}$Kavli Institute for Cosmology, Madingley Road, Cambridge, CB3 0HA, UK\\
$^{3}$Institute of Astronomy, University of Cambridge, Madingley Road, Cambridge, CB3 0HA, UK\\
$^{4}$DAMTP, Centre for Mathematical Sciences, University of Cambridge, Wilberforce Road, Cambridge, CB3 OWA, UK\\
$^{5}$Raman Research Institute, C V Raman Avenue, Sadashivanagar, Bangalore 560080, India\\
$^{6}$School of Physics and Astronomy, Tel-Aviv University, Tel-Aviv 69978, Israel
}

\date{Accepted XXX. Received YYY; in original form ZZZ}

\pubyear{2023}

\begin{document}
\label{firstpage}
\pagerange{\pageref{firstpage}--\pageref{lastpage}}
\maketitle

\begin{abstract}
The early Universe, spanning 400,000 to 400 million years after the Big Bang ($z\approx1100-11$), has been left largely unexplored as the light from luminous objects is too faint to be observed directly. 
While new experiments are pushing the redshift limit of direct observations,  measurements in the low-frequency radio band promise to probe early star and black hole formation via observations of the hydrogen 21-cm line. 
In this work we explore synergies between 21-cm data from the HERA and SARAS 3 experiments and observations of the unresolved radio and X-ray backgrounds using multi-wavelength Bayesian analysis. We use the combined data set to constrain properties of Population II and Population III stars as well as early X-ray and radio sources. 
The joint fit reveals a 68 percentile disfavouring of Population III star formation efficiencies $\gtrsim5.7\%$. We also show how the 21-cm and the X-ray background data synergistically constrain opposite ends of the X-ray efficiency prior distribution to produce a peak in the 1D posterior of the X-ray luminosity per star formation rate. We find (at 68\% confidence) that early galaxies were likely 0.3 to 318 times as X-ray efficient as present-day starburst galaxies. 
We also show that the functional posteriors from our joint fit rule out global 21-cm signals deeper than $\lesssim-203\ \mathrm{mK}$ and power spectrum amplitudes at $k=0.34\ h\mathrm{Mpc^{-1}}$ greater than $\Delta_{21}^2 \gtrsim 946\ \mathrm{mK}^2$ with $3\sigma$ confidence.
\end{abstract}

\begin{keywords}
stars: Population III -- early Universe -- dark ages, reionization, first stars -- X-rays: diffuse background
\end{keywords}



\section{Introduction} \label{sec:intro}

Over the last few decades, the early Universe was studied through various probes across the electromagnetic spectrum. 
Through its deep surveys in legacy fields \citep{hubble_hudf} and detection of galaxies at redshifts as high as $z\approx 11$ \citep{hubble_gnz11}, the \textit{Hubble Space Telescope} was vital in contributing to our understanding of the Universe.  
Its successor, the \textit{James Webb Space Telescope} \citep[\textit{JWST},][]{jwst}, is expected to push this limit even further by undertaking even deeper surveys \citep[e.g. ][]{jwst_glass,jwst_ceers,jwst_jades_nircam,jwst_jades_nirspec}. High-redshift candidates have already been reported at $z\approx20$ \citep{jwst_z20_candidates} and spectroscopically confirmed galaxies have been detected at $z\approx 13$ \citep{jwst_z13,jwst_z13_2}. Complementary to galaxy surveys, observations of the Cosmic Microwave Background (CMB) at redshift $z\sim 1100$ with probes such as \textit{COBE} \citep{cobe}, \textit{WMAP} \citep{wmap}, ACT \citep{act}, and the \textit{Planck} mission \citep{planck2013_overview,planck2015_overview,planck2018_overview} were fundamental for our understanding of large scale cosmology.  
Despite these multiple observational successes, there remains a gap in our 
observations of the early Universe. Epochs including the cosmic Dark Ages, Cosmic Dawn, and the start of the Epoch of Reionization remain largely uncharted. Despite the growing observational capabilities, the first luminous objects will likely remain beyond reach, being either too faint or obscured by the neutral intergalactic medium (IGM) to be detected directly even by state-of-the-art space telescopes like \textit{JWST}. 

The 21-cm line of neutral hydrogen provides a promising probe of these unexplored epochs \citep[for reviews]{furlanetto_review,pritchard_review,barkana_review, mesinger_review}. 
The strength of the 21-cm signal is governed by a range of physical processes that modulate the contrast between the spin temperature and background radiation temperature.  
Hence the radiation from the first stars and black holes, that formed from the collapsed dark matter overdensities, has a measureable impact on the 21-cm signal \citep{yajima2015, mirocha_lradio, schauer2019_popIII_21cm, reis2021subtlety, reis2022shot, munoz2022}. 
Consequently, we can use the 21-cm signal to infer the properties of the first objects that illuminated the early Universe, and constrain a range of features related to the first stars like the mass distribution, spectral emissivity, and star formation efficiency \citep{schauer2019_popIII_21cm,thomas_popIIIimf}. Stars may also form X-ray binaries, that heat the IGM, and produce an observable signature in the 21-cm signal, that can be used to constrain the X-ray efficiency and X-ray spectral energy distribution (SED) \citep{Pacucci_2014,fialkov2014observable, fialkov_fstarfX}. In death, the remnants of the first stars also affect the IGM as supernovae, e.g. by X-ray production from inverse Compton scattering or bremsstrahlung \citep{mirocha2018_popIII_21cm}, or at longer radio wavelengths through synchrotron radiation \citep{jana2019_SNe}. These radio sources contribute to a galactic inhomogenous radio background which enhances the contrast between the spin temperature and the background radiation temperature \citep{reis2020} allowing us to constrain properties like the radio efficiency of high redshift galaxies. The 21-cm line may also inform us about other sources contributing to the radio background such as primordial black holes \citep{mittal2022_pbh}, and it has also been considered in constraining homogenous radio backgrounds \citep{AF_2019_rad, thomas2024_cosmic_strings} arising from exotic models of interacting dark matter \citep{barkana_dm_interact,munoz2018_dm_interact,fraser2018_dm,liu2019_dm,jones2021_dm,hibbard2022_dm} and superconducting cosmic strings \citep{brandenberger_cosmic_strings,cyr_cosmic_strings,thomas2024_cosmic_strings}. 
The 21-cm line is also the only probe of the Dark Ages, which is free from astrophysical sources and strictly governed by gravity \citep{lewis2007_DA, mondal2023_DA,fialkov2023_cosmic_mysteries}, and can therefore probe the initial conditions of the Universe. The 21-cm angular bispectrum could be used as a probe of inflationary physics by detecting non-Gaussianities from non-linear collapse \citep{Pillepich2007_nonGaussian}.

While the 21-cm line has great potential to inform us about physical processes in the early Universe, the detection of the signal poses numerous challenges. Galactic synchrotron radiation and extragalactic contributions from e.g. radio galaxies and free-free emission make up a significant source of contamination as they are several orders of magnitude brighter than the cosmological signal \citep{dimatteo_foregrounds,dimatteo_foregrounds2,oh_mack_foregrounds, cooray_foregrounds,jelic_foregrounds}.

There are two types of experiments attempting to observe the 21-cm signal. Single antenna radio experiments that measure the sky-averaged (global) 21-cm signal, and radio interferometer arrays which record fluctuations of the radio sky thus targeting the 21-cm power spectrum. Both types of experiments require careful calibration and error estimation to account for the galactic foregrounds and intrinsic systematics. 

At the moment of writing, there has only been one tentative detection of the global 21-cm signal reported by the EDGES collaboration \citep{EDGES_detection}. In their study \citet{EDGES_detection} find a deeper-than-expected absorption trough that is best modelled by a flattened Gaussian, and which requires non-standard astrophysical or cosmological models to be explained as the global 21-cm signal \citep{barkana_dm_interact,munoz_dm_interact,ewall-wice_bh_explain, Feng_18}. Currently unconfirmed, the EDGES detection could be an artifact of the data analysis, e.g. a potential signature of instrumental systematics \citep{Hills_EDGES_concerns, saras_EDGES_concerns,Sims_EDGES_concerns, bevins_maxsmooth}, a speculation that is consistent with the null detection by the SARAS~3 global signal experiment \citep{saras3, saras3_measurement}. The field is very active with several other global signal experiments like LEDA \citep{leda}, REACH \citep{reach}, MIST \citep{mist}, and PRIZM \citep{prizm} also attempting to detect the 21-cm global signal. 

Although no detections of the 21-cm power spectrum have been made, upper limits have been reported by experiments such as LOFAR \citep{lofar_paper, lofar_limits, ACE}, PAPER \citep{paper_paper, paper_limits}, LEDA \citep{leda_ps}, MWA \citep{mwa_paper,mwa_limits}, HERA \citep{HERA,HERA_IDR2a_limits,HERA_IDR3} and NenuFAR \citep{nenufar}. These upper limits will be lowered as experiments improve, before a potential detection with experiments such as HERA, LOFAR, or the SKA1-LOW \citep{ska_paper}. 

In this work, we conduct a multi-wavelength study combining current 21-cm observations with observations of the unresolved X-ray and radio backgrounds to constrain properties of the early Universe. Specifically, we compare early Universe simulations with 
the upper limits on the 21-cm power spectrum from HERA \citep[second public release of data from the H1C observing season,][]{HERA_IDR3}, 
the limits on the global 21-cm signal derived using the SARAS~3 data \citep{saras3_measurement}, 
collated measurements of the present-day radio background temperature \citep{tradio_data}, and collated measurements of the Cosmic X-ray background (CXB) from \citet{harrison2016} and \citet{hickox2006}. 
This combined data set allows us to put joint constraints on the latest set of 21-cm models created with the code 21-cm Semi-numerical Predictions Across Cosmic Epochs (now called \textsc{21cmSPACE}) which, in addition to the previously incorporated effects  
\citep[e.g.][]{visbal2012signature,fialkov2014observable}
now includes a separate star formation prescription for Population II (Pop II) and Population III (Pop III) stars \citep{popIII_implement, thomas_popIIIimf} and line-of-sight radio fluctuations \citep{los_radio_fluc}. This latest development allows us to constrain the two stellar populations separately and ensures the X-ray limits are not overestimated \citep{PopIII_weakens_Xray_bounds}. The models considered here also include an excess radio background in addition to the CMB, originally motivated by the work of \citet{Feng_18} and the EDGES detection \citep{EDGES_detection}. We assume that the excess radio background is created by high-redshift galaxies \citep{reis2020} and model line-of-sight radio fluctuations \citep{los_radio_fluc}.

To perform our data analysis \textsc{21cmSPACE} is used to generate realisations of the Universe, and emulators are trained on the simulation outputs, providing evaluations of expected observables in fractions of a second. This allows us to conduct an exploration of the parameter space using nested sampling \citep{ns_skilling, ns_ashton}  in a reasonable time and to infer constraints on astrophysical parameters. 
In this work multiple nested sampling runs are conducted, one with each individual observational constraint, and one with all of them jointly. We find that our joint analysis approach provides the strongest constraints on our astrophysical parameters as we are able to exploit the synergies and leverage the strengths of each observational data set. 
While combined constraints have already been considered in \citet{HERA_IDR2b_theory} using the upper limits on the 21-cm power spectrum, X-ray and radio backgrounds, it was done in a schematic post-processing step. Here we include the data constraints directly in the likelihood during sampling. We also use the improved upper limits on the 21-cm power spectrum \citep{HERA_IDR3}. A similar joint analysis was carried out by \citet{bevins_joint_paper}, however only the upper limits on the 21-cm power spectrum and global signal residuals were included in the likelihood. Thus, this work constitutes the first study to include all these multi-wavelength observations jointly to constrain properties of the early Universe. 

In Section \ref{sec:theory} we describe the early Universe simulation code \textsc{21cmSPACE}. In Section \ref{sec:data} we introduce the observational data sets, and we establish the methodology used to train emulators for parameter inference. In Section \ref{sec:results} we present the astrophysical parameter constraints from our analysis and show functional posteriors on the 21-cm global signal and 21-cm power spectrum. In Section \ref{sec:discussion} we discuss potential improvements to our analysis, and, lastly, we conclude this study in Section \ref{sec:conclusion} by summarising the findings from our analysis and highlighting some of the key takeaways.
\section{Simulating the Early Universe} \label{sec:theory}
To constrain high-redshift astrophysics, we require a model of how astrophysical processes impact observables. 
We thus begin by describing the semi-numerical simulation code \textsc{21cmSPACE}.
While parameter inference studies were conducted with previous iterations of these 21-cm models \citep[see e.g.][]{HERA_IDR2b_theory, bevins_saras3, bevins_joint_paper}, we use an updated version of \textsc{21cmSPACE}, which now includes self-consistent calculation of the high redshift contributions to the present-day radio background temperature and the integrated X-ray Background, as well as improved astrophysical modelling. 

\subsection{Overview of 21cmSPACE}
\label{sec:simulations}
In \textsc{21cmSPACE} the early Universe is modelled as a periodic box split into cubic cells. 
For our simulations, this box is composed of $128^3$ cells, each with a side length of $3$\,comoving Megaparsecs (cMpc). 
As a result, the total simulation box covers a cosmological volume of $(384\ \mathrm{cMpc})^3$.
These large volumes are necessary to ensure a good statistical sample of observables, in particular for the 21-cm global signal and 21-cm power spectrum on large scales, where the HERA upper limits are strongest~\citep{HERA_IDR3}.
However, if simulations are to run in computationally feasible time frames, this cosmological size comes at the cost of requiring large cell sizes, $3$\,cMpc. 
The simulation code thus adopts a semi-numerical approach to modelling the early Universe.
Phenomenology, on scales larger than the cell size processes such as radiative transfer~\citep[e.g.][]{reis2020} and bulk baryon dark-matter relative velocities~\citep{Tseliakhovich_2010, visbal2012signature, Fialkov_2012} are modelled numerically, while physics on smaller scales, like star formation~\citep[e.g.][]{popIII_implement} and halo growth~\citep[e.g.][]{Barkana_2004}, are treated using analytic models or fitting formulas.
Using this approach, each simulation of the early Universe takes a few hours.

A simulation is initialized at $z=50$ (towards the end of the cosmic dark ages) with a set of cosmological initial conditions for matter overdensity, gas kinetic temperature, ionization fraction, and baryon dark-matter relative velocities generated using \textsc{CAMB}~\citep{Lewis_1999, Lewis_2002, Lewis_2011} and \textsc{RECFAST}~\citep{RECFAST}, assuming a \textit{Planck 2018} best-fit $\Lambda$CDM cosmology~\citep{planck2018}.
These large-scale fields are then evolved forward at each time step.
Large-scale overdensity and relative velocities fields are analytically propagated forward via linear perturbation theory~\citep[see e.g.][]{barkana_review}. 
Whereas, the gas kinetic temperature is evolved in each cell by integrating the heating differential equation, taking into account adiabatic cooling, ionization cooling, structure formation heating, and heating or cooling from astrophysical sources~\citep[see][for the most recent summary of the code]{TG_CRs2023}. 
Lastly, the ionization fraction is computed in a two-stage approach, with an excursion set formalism~\citep{Furlanetto_2004} used to model large-scale fully ionized bubbles, and solving the ionization differential equation used to model partial ionization outside the bubbles~\citep{Mesinger_2013}, e.g. due to X-rays \citep{fialkov_fstarfX}. 

The evolution of the gas temperature and ionization fraction thus requires modelling of astrophysical sources and their radiation fields. 
First, the distribution of dark matter halos within each cell is found analytically using a hybrid mass function~\citep{Barkana_2004, Fialkov_2012}.
Then star-forming halos are identified via a minimum mass threshold for star formation, considering the impact of the Lyman-Werner~\citep{Fialkov_2013}, baryon-dark matter velocity~\citep{Fialkov_2012}, and photoheating feedback effects~\citep{Cohen_2016}. 
The star formation history of these halos then follows the prescription of \citet{popIII_implement}.
The halos first form a burst of metal-free Population III stars. 
Supernovae from these stars then enrich the halo with metals.
However, the same supernovae also suppress star formation by heating and ejecting material from the halo, there is thus a recovery time for the halo, after which it starts quiescently forming metal-containing Population II stars. 
With the star formation rate of each cell, the emissivity of various radiative species can be calculated, for example, in the Lyman band, ionizing UV, radio, and X-ray. 
We model the propagation of the Lyman band~\citep{AF2014_lya, reis2021subtlety}, radio~\citep{reis2020, los_radio_fluc}, and X-ray photons~\citep{fialkov2014observable, fialkov_fstarfX} through the simulation cube using window functions, taking into account the lightcone effect and the redshifting of the source spectrum as the Universe expands. 
Ionizing UV propagation is instead modelled by the earlier mentioned excursion set formalism. 
With this radiative transfer prescription, a closed set of equations can be formed, allowing the Universe to be propagated forward from $z = 50$, in our case ending at $z \sim 6$ when the 21-cm signal is expected to be extinguished by reionization \citep{fan2006_end_eor,mcgreer2014_xHI,planck2018, 2018Natur.553..473B, 2023ApJ...942...59J}.
Ultimately, the simulation calculates cubes of star formation rate, gas temperature, ionization fraction, and other quantities which are then used to compute our observables of interest.

\subsection{Simulation of observables}~\label{sssec:obs}
The 21-cm signal is produced by the neutral atomic hydrogen that permeates the Universe before reionization~\citep[for review articles see,][]{furlanetto_review,pritchard_review,barkana_review, mesinger_review}. 
Hydrogen atoms have a hyperfine transition with energy $E_{\rm 10} = 5.87$\,$\mu$eV corresponding to a wavelength of 21\,cm.
As a result, hydrogen can emit or absorb at this line, enhancing or diminishing a radio background at rest-frame 21\,cm wavelength, creating the 21-cm signal. 
As this hyperfine transition is very narrow, photons rapidly redshift out of the line, consequently, different epochs can be probed through the net emission or absorption seen today at different radio frequencies~\citep{Madau_1997}.

The strength of this signal is hence dependent on the number density of neutral hydrogen in the early Universe, the strength of the hyperfine transition line, and the relative population of the hyperfine states of neutral hydrogen.
It is common to quantify the latter of these in terms of the statistical spin temperature $T_{\rm S}$~\citep{Scott_1990}
\begin{equation}
    \frac{n_{\rm 1}}{n_{\rm 0}} = 3 \exp\left(- \frac{E_{\rm 10}}{k_{\rm B} T_{\rm S}} \right),
\end{equation}
where $n_1$ and $n_0$ are the number densities of neutral hydrogen atoms in the hyperfine states where the magnetic moments between the proton and electron are aligned and anti-aligned respectively, and the factor of 3 is the relative statistical weight of the singlet and degenerate triplet states. 
$T_{\rm S}$ in turn can be computed using the coupling equation~\citep{Field_1958,furlanetto_review,Venumadhav_2018}, 
\begin{equation}
    \frac{1}{T_S} = \frac{x_c T_\mathrm{K}^{-1} + x_\gamma T_\gamma^{-1} + x_\alpha T_\alpha^{-1}}{x_c + x_\gamma + x_\alpha}.
\end{equation}
Here $T_\mathrm{K}$ is the kinetic gas temperature, $T_\gamma$ is the background radiation temperature, $T_\alpha$ is the colour temperature of the Ly\,$\alpha$ radiation field, and $x_c$, $x_\gamma$, and $x_\alpha$ are the corresponding coupling coefficients to these temperatures due to collisional coupling, radiative coupling, and the Wouthuysen-Field effect~\citep{Wouthuysen_1952, Field_1958} respectively.
Similarly, the number density of neutral hydrogen and the strength of the hyperfine transition can be condensed into the optical depth of the 21-cm line
\begin{equation}
\tau_{\rm 21} = \frac{3}{32 \pi} \frac{h c^3 A_{\rm 10}}{k_{\rm B} \nu_{\rm 21}^2} \left[\frac{x_{\rm HI} n_{\rm{H}}}{(1+z)^2 (d{\rm v}_\parallel/dr_\parallel)} \right]\frac{1}{T_{\rm S}},
\label{eqn:21cm_optical_depth}
\end{equation}
where $\nu_{\rm 21} = 1420$\,MHz is the frequency of the 21-cm line, \mbox{$A_{\rm 10} = 2.85 \times 10^{-15}$\,s$^{-1}$} the spontaneous emission rate of the 21-cm transition, $x_{\rm HI}$ the neutral atomic hydrogen fraction, $n_{\rm H}$ the number density of all hydrogen, and $d{\rm v}_\parallel/dr_\parallel$ the proper velocity gradient parallel to the line of sight.
Combining the above, the magnitude of the 21-cm signal from a given redshift can then be expressed as a differential brightness temperature seen today using the equation of radiative transfer
\begin{equation}
    T_{\rm 21} = \left(1-e^{-\tau_{\rm 21}} \right) \frac{T_{\rm S} - T_\gamma}{1+z}.
\end{equation}
$T_\gamma$ is often assumed to be the CMB temperature.
However, in this study, we include an excess radio background from high-redshift galaxies~\citep{reis2020}, motivated by the measurements of the 
low-frequency radio emission in galaxies \citep{gurkan_lradio} as well as by the reports of an excess radio background by ARCADE2 \citep{arcade2, arcade2_seiffert} and LWA1 \citep{tradio_data}.
The radiation temperature is then the sum of the CMB temperature and this excess.  

\textsc{21cmSPACE} solves the above equations for us, outputting cubes of $T_{\rm 21}$ at different redshifts. 
However, current 21-cm observations are not trying to make maps of the 21-cm signal, but rather probe summary statistics of the field that are easier to measure~\citep{lofar_paper, mwa_paper, HERA, prizm, reach, mist}. 
We, hence, compress the output of \textsc{21cmSPACE} into the 21-cm global signal $\langle T_{\rm 21} \rangle$, the average of $T_{\rm 21}$ at a given redshift, and the 21-cm power spectrum $P(k, z)$, i.e. the Fourier transform of the two-point correlation function of $T_{\rm 21}$ at redshift $z$ and comoving wavenumber $k$ (in units of $h\,\mathrm{Mpc}^{-1}$, where $h$ is the normalized Hubble parameter $h=0.674$). These can then be compared to observational constraints as part of our joint analysis.
Since HERA express their 21-cm power spectrum upper limits~\citep{HERA_IDR2a_limits, HERA_IDR3} using the convention
\begin{equation}
    \Delta_{\rm 21}^2 (k, z) = \frac{k^3}{2\pi^2} P(k, z),
\end{equation}
we also adopt this form of the 21-cm power spectrum throughout this study, where $\Delta_{\rm 21}^2$ has units of mK$^2$.

Internally, \textsc{21cmSPACE} models an excess radio background from high redshift galaxies \citep{reis2020} following the methodology introduced by  \citet{tradio_equation}. For self-consistency, we use the same prescription to compute the present-day excess radio background from high redshift sources.
Galaxies are modelled as having a luminosity per unit frequency proportional to their star formation rate SFR~\citep[as is seen in observations of radio galaxies,][]{gurkan_lradio,mirocha_lradio},
\begin{equation}~\label{eqn:radio_luminosity}
    L_\mathrm{r}(\nu) = f_{\rm r} 10^{22} \left( \frac{\nu}{\rm 150\,MHz} \right)^{-0.7} \frac{\rm SFR}{\rm M_{\odot} yr^{-1}},
\end{equation}
where $f_{\rm r}$ is the radio emission efficiency relative to present-day galaxies ($f_{\rm r} = 1$), and the spectral index of 0.7 is chosen to match the observed spectrum of low-redshift radio galaxies \citep{hardcastle_lradio_specind,gurkan_lradio}. 
In the version of the code we use for this study, $f_{\rm r}$ is assumed to be the same for Pop~II and Pop~III galaxies. 
From equation~\eqref{eqn:radio_luminosity} and the star formation rate density output by \textsc{21cmSPACE}, the comoving luminosity density per unit frequency $\epsilon_{\rm r}(\nu, z)$ can be calculated.
The sky-averaged excess radio background today from high redshift ($z > 6$) sources is then given by integrating over emission redshifts $z'$ 
\begin{equation}\label{eq:trad_equation}
    T_{\rm r}(\nu,z = 0) =  \frac{c^3}{8 \pi k_{\rm B} \nu_{\rm 21}^2}  
    \int_{z' = 6}^{\infty} \epsilon_r \left[ \nu (1 + z'), z' \right] \frac{(1+z')}{H(z')} dz',
\end{equation}
where $H$ is the Hubble parameter.
We perform this calculation in post-processing similar to \citet{reis2020}. As the simulations end at $z=6$ contributions to the radio background from sources $z<6$ are not included. Excluding lower redshift sources makes our constraints more conservative as the limits would only get stronger with the inclusion of $z<6$ sources.

The unresolved X-ray background from high redshift sources has previously been calculated by post-processing \textsc{21cmSPACE} outputs~\citep{fialkov_fstarfX, HERA_IDR2b_theory, HERA_IDR3} in a similar manner to which we compute the excess radio background.
However, for this study, we integrate the code to perform this calculation into \textsc{21cmSPACE} to have this as a standard output.
Within \textsc{21cmSPACE}, X-ray sources (both Pop~II and Pop~III) are assumed to have an X-ray luminosity that follows a present-day starburst galaxy like relation $L_{\rm X} /{\rm SFR} = 3\times 10^{40} f_{\rm X}$\,erg\,s$^{-1}$\,M$_{\odot}^{-1}$\,yr~\citep{lx_sfr_grimm, lx_sfr_ranalli, lx_sfr_gilfanov, lx_sfr_furlanetto,lx_sfr_mineo}, where $f_{\rm X}$ is the X-ray emission efficiency of high redshift sources (both Pop~II and Pop~III) normalized to the luminosity predicted by \citet{fragos_fx1_2}.
The X-ray SED is similarly parameterized in the code due to the uncertainties in the astrophysics of the early Universe. 
In this study, we consider X-ray SEDs with a lower cutoff energy $E_{\rm min}$ and power-law slope $\alpha$, though our X-ray background code is capable of handling any SED.
By combining the parameterized X-ray SED, X-ray efficiency, and the star formation rate density already calculated in \textsc{21cmSPACE}, we can compute the $\epsilon_{\rm X}(z, E)$, specific X-ray emissivity, throughout the simulation. 
From this, we compute the present-day X-ray specific intensity from high redshift ($z \geq 6$) sources via~\citep[e.g.][]{pritchard_xrb}
\begin{equation}~\label{eq:xraybackground} 
    J_{\rm X}(E) = \frac{c}{4 \pi} \int_{z' = 6}^{\infty} \epsilon_{\rm X} \left[ E (1 + z'), z' \right]  \frac{e^{- \tau_{\rm X}(z', E)}}{(1 + z') H(z')} dz',
\end{equation}
where $\tau_{\rm X}(z', E)$ is the optical depth of X-rays between their emission redshift of $z'$ and the present day. 
$\tau_{\rm X}$ itself is calculated by integrating over the weighted X-ray cross-section of species~\citep{Verner_1996} in the IGM between $z'$ and the present-day. We assume H\textsc{\,i}, He\textsc{\,i}, and He\textsc{\,ii} dominate $\tau_{\rm X}$ given their much greater abundance than metals in the IGM. 
This calculation is performed self-consistently with the changing abundances of these species within the \textsc{21cmSPACE} simulation due to ionization.
The computed $J_{\rm X}$ is output by the simulation from $0.2$\,keV to $95.0$\,keV, and can be later compared to observations to constrain high-redshift astrophysics. As with the radio background temperature, X-ray contributions are only included down to $z=6$ where the simulations end. Following a similar argument, this constitutes a conservative approach where any astrophysical limits from the X-ray data would only become stronger if lower redshift constributions were to be included.

\subsection{21cmSPACE parameters and settings}
As alluded to in the previous section, \textsc{21cmSPACE} takes several input parameters that describe the uncertain astrophysics and cosmology of the early Universe. 
For clarity, we now outline which of these parameters we vary in this study. 
Additionally, we state the priors we use on these parameter values as part of our constraints: 
\begin{itemize}
    \item $f_\mathrm{\ast,III}$ - Efficiency of Pop~III star formation. Sampled from a log-uniform prior $f_\mathrm{\ast,III} \in [10^{-3}, 10^{-0.3}]$.
    \item $f_\mathrm{\ast,II}$ - Efficiency of Pop~II stars formation. Similarly, sampled from a log-uniform prior $f_\mathrm{\ast,II} \in [10^{-3}, 10^{-0.3}]$. 
    \item $t_\mathrm{delay}$ - The recovery time of a star-forming halo between the first Pop~III supernovae and the beginning of Pop~II star formation. We sample time-delay values from the discrete uniform prior \mbox{$t_\mathrm{delay} \in \{10, 30, 100\} \mathrm{\,Myr}$}.
    \item $V_{\rm c}$ - Minimum circular virial velocity of a halo for star formation (in the absence of feedback). If no feedback effects are present, gas collapse into stars occurs when the halo reaches a mass that makes the virial temperature exceed $T_\mathrm{crit}=7300\mathrm{\,K} (V_{\rm c}/10\ \mathrm{km\,s^{-1}})^2$ \citep{vc_crit_sf}. The Lyman-Werner, baryon-dark matter velocity, and photoheating feedbacks (all included in our simulations) increase this mass threshold. We consider a range of star formation thresholds by sampling $V_c$ from a log-uniform prior $V_c \in [4.2, 100]\mathrm{\,km}\,\mathrm{s^{-1}}$. In this range $V_c<16.5$ corresponds to molecular cooling, and $V_c \geq 16.5$ corresponds to atomic cooling. As Lyman-Werner feedback is only relevant for molecular cooling star formation, it is automatically disabled in the code once the threshold for atomic cooling is passed. 
    \item $f_{\rm X}$ - The relative X-ray emission efficiency of early galaxies, with $f_{\rm X} = 1$ corresponding to the prediction for low metallicity X-ray binaries from \citet{fragos_fx1} and \citet{fragos_fx1_2}. See the previous section for more details. $f_{\rm X}$ is sampled from a log-uniform prior $f_{\rm X} \in [10^{-3}, 10^3]$.
    \item $\alpha$ - Power-law slope of the X-ray SED. See the previous section for more details. Sampled from a discrete uniform prior, $\alpha \in \{1, 1.3, 1.5\}$.
    \item $E_\mathrm{min}$ - Lower-energy cutoff of the X-ray SED (eV). See the previous section for more details. Sampled from a discrete uniform prior, $E_\mathrm{min} \in \{100,200,300,400,500,600,700,800,900,1000,1100,\\1200,1300,1400,1500,2000,3000\}\textrm{\,eV}$. 
    \item $\zeta$ - Effective ionization efficiency of galaxies. Rather than using $\zeta$ directly as part of our constraints, we instead use $\tau$, the optical depth to reionization~\citep[as has been previously done, e.g. in][]{cohen2017, simcode4_cohenemu, bevins_saras3}. $\tau$ is an output of \textsc{21cmSPACE} rather than an input, but, as it monotonically increases with $\zeta$, it can be used in place of $\zeta$ for the training of our emulators, and thus we can constrain it directly in our analysis. We adopt a uniform prior with $3\sigma$ around the measured value from the \citet[$\tau=0.054 \pm 0.007$]{planck2018}, e.g. $\tau \in [0.033, 0.075]$.
    \item $f_{\rm r}$ - Relative radio emission efficiency of early galaxies, defined via the observational $L_\mathrm{r}-\mathrm{SFR}$ relationship \citep[][equation~\eqref{eqn:radio_luminosity}]{gurkan_lradio,mirocha_lradio}, with $f_{\rm r}=1$ corresponding to the radio efficiency of present-day galaxies. Throughout this paper, any quoted $L_\mathrm{r}/\mathrm{SFR}$ value is evaluated at the $\nu=150\mathrm{\,MHz}$ reference frequency and with a spectral index of $0.7$ which is compatible with observations \citep{hardcastle_lradio_specind, gurkan_lradio}. To explore a broad range of potential astrophysics, radio efficiencies are sampled from a log-uniform prior, $f_r \in [10^{-1}, 10^{5}]$.
\end{itemize}
\textsc{21cmSPACE} has other parameters we kept fixed, due to them only having weak effects on the 21-cm signal for the purposes of this study.
These were:
\begin{itemize}
    \item $R_\mathrm{mfp}$ - The maximum mean free path of ionizing photons. Fixed to, 40\,cMpc~\citep{Wyithe_2004}.
    \item Signature of Pop III IMF (the initial mass distribution of Population III stars) in the Lyman band radiation \citep{thomas_popIIIimf}. In this study, the IMF is fixed to being logarithmically uniform in the mass range $2-180$\,M$_\odot$~\citep{thomas_popIIIimf, Klessen_2023}.
\end{itemize}
Furthermore, there are various settings in \textsc{21cmSPACE}, representing the modelling of different physics, which can be enabled and disabled. We set these to:
\begin{itemize}
    \item Baryon dark-matter relative velocities \citep{Fialkov_2012, visbal2012signature} enabled.
    \item Lyman-Werner feedback \citep{Fialkov_2013} and photoheating feedback \citep{Cohen_2016} enabled. 
    \item X-ray heating \citep{visbal2012signature, fialkov2014observable, fialkov_fstarfX}, CMB heating~\citep{AF_2019_rad}, and Ly\,$\alpha$ heating~\citep{reis2021subtlety} enabled.
    \item Ly\,$\alpha$ multiple scattering~\citep{reis2021subtlety} enabled.
    \item Star formation efficiency suppression~\citep{Fialkov_2013} enabled. 
    \item Poisson fluctuation of galaxy formation and star formation efficiency~\citep{reis2020} is disabled as fluctuations are negligible to the power spectrum in the low redshift HERA bands relevant to this paper \citep{reis2022shot}. 
\end{itemize}
\subsection{Recent updates to 21cmSPACE}
Early Universe models generated from an older version of \textsc{21cmSPACE} were previously constrained using the data of SARAS~2, SARAS~3, EDGES High Band, LOFAR, MWA and HERA \citep{SARAS2_2017, SARAS2_2018, EDGESHI_2019, LOFAR_2020, BevinsSARAS2_2022, HERA_IDR2b_theory,HERA_IDR3,bevins_saras3, bevins_joint_paper} as well as trying to understand the tentative EDGES Low Band detection \citep[e.g.][]{AF_2019_rad, reis2020}. 
However, the code has undergone several updates since then. 
For ease of comparison and the convenience of those familiar with our previous works, we summarise here the relevant code improvements.

The code previously modelled either Pop II star or Pop III star formation in one simulation, but not both.
\textsc{21cmSPACE} can now model both stellar populations in the same simulation and the transition between them using the star formation prescription of \citet{popIII_implement}. The Pop III star formation prescription is implemented in \textsc{21cmSPACE} using fitting formulae derived from \textsc{ASLOTH} simulations~\citep{ASLOTH}. 
In this prescription, when a halo reaches the critical mass for star formation (set by a combination of $V_{\rm c}$ and feedback effects), a burst of Pop III star formation occurs.
Most of these Pop III stars are expected to be short-lived (with the exact fraction depending on the Pop III IMF), and so it is assumed that the first Pop III supernovae go off well before a second round of Pop III star formation can occur. These violent explosions enrich the star-forming halo with metals and disrupt it, heating and ejecting its gaseous contents. It takes some time before the enriched gas re-collapses into the halos and becomes available for the next round of star formation. This process is modelled as a time delay, $t_{\rm delay}$, between the supernovae and the onset of Pop II star formation. 
Since Pop III star formation and Pop II star formation are now treated separately, we assign different star formation efficiencies, $f_{\rm *,II}$ and $f_{\rm *,III}$ (replacing the previously used $f_*$ parameter). The inclusion of Pop~III star formation allows us to potentially constrain Pop III galaxy properties using the observables we consider in the following section. Our earlier SARAS~3 analysis \citep{bevins_saras3} was done on the previous set of models which did not differentiate between Pop II and Pop III. 
Note that, accounting for the contribution from Pop III stars, separate from Pop II stars, has been shown in \citep{PopIII_weakens_Xray_bounds} to be necessary to avoid overstating X-ray emissivity constraints. 

In addition to the separation of Pop II and Pop III stars, the upgraded version of \textsc{21cmSPACE} includes the improved modelling of Pop III stars introduced in \citet{thomas_popIIIimf}. Compared to the previous studies, the Lyman band and Lyman-Werner band spectra are now derived self-consistently given an IMF using a set of individual stellar metal-free spectra. 
The finite lifetime of Pop III stars, which can impact the 21-cm signal at high redshifts, is also modelled, though it is unlikely to have much of an effect in this study for the chosen IMF. The Lyman band and Lyman-Werner band spectra for Pop II stars remain fixed~\citep{Starburst_1999}.
In our version of the code, the X-ray, radio, and ionizing emission efficiencies of Pop~II and Pop~III star galaxies are still assumed to be the same, and hence there are only single $f_{\rm X}$, $f_{\rm r}$, and $\zeta$ parameters. While a separation of the X-ray spectra by stellar population is underway (Gessey-Jones et al. in preparation), modelling separate radio efficiencies remains to be implemented. We consider the assumption of a simple model with identical radio efficiencies for Pop~II and Pop~III stars, e.g. $f_\mathrm{r}=f_\mathrm{r,II}=f_\mathrm{r,III}$, a safe zero-order approximation. 
As the Pop III radio emission would be proportional to a separate radio efficiency and SFR (equation~\eqref{eqn:radio_luminosity}), the relative Pop III radio background contribution would be largest at high redshifts where the Pop III SFR is larger than, or more competitive with, the Pop II SFR. Thus, high redshift experiments might be able to constrain the Pop III radio efficiency by measuring the signature of Pop III radio emission on the 21-cm signal. However, this high redshift signature will be suppressed, as the higher CMB temperature requires an even stronger radio background to impact the 21-cm signal, compared to low redshifts. Constraints from experiments measuring the cumulative present-day radio background temperature are unlikely to change as Pop II stars dominate the radio background from low to intermediate redshifts, and radio contributions from high redshift sources are diminished due to redshifting. Therefore, we would not expect the constraints, from the data used in this study, to change significantly. However, improved data from future high redshift experiments will allow us to constrain properties like the radio efficiency of Pop III stars. We leave the development of a robust model with separate radio efficiencies in \textsc{21cmSPACE} to future work. 

Models with excess radio background from high redshift galaxies~\citep{reis2020} have been used in our prior constraints \citep[e.g.][]{bevins_saras3, HERA_IDR3}. However, we used an approximate treatment of this effect, assuming that each hydrogen cloud experienced an isotropic radio background averaged over sightlines.
This is contrary to reality since the observed 21-cm emission or absorption is actually along the line of sight.
The newly upgraded version of \textsc{21cmSPACE} used in this study, includes this line-of-sight effect for the excess radio background, following its implementation and investigation in \citet{los_radio_fluc}.
In this original study, it was found that in extreme radio excess models, the inclusion of line-of-sight radio fluctuations might cause up to two orders of magnitude difference in the 21-cm power spectrum and 5\% difference in the global signal depending on the values of astrophysical parameters.
Hence, it is an important effect for us to include for reliable constraints on $f_{\rm r}$ and parameters degenerate with $f_{\rm r}$.

Finally, for this paper, we have added a module to \textsc{21cmSPACE} to automatically calculate the contribution to the unresolved X-ray background from high redshift sources. We described this calculation in more detail in section~\ref{sssec:obs}.
\section{Data and parameter inference} \label{sec:data}
We now move on to describing the data sets we will use to constrain the \textsc{21cmSPACE} models discussed in the last section. In subsections~\ref{sec:h1c_idr2}, \ref{ssec:global}, \ref{ssec:xrb}, and \ref{ssec:erb}, we describe existing observational constraints on the 21-cm power spectrum, 21-cm global signal, unresolved X-ray background, and excess radio background that we include in our analysis. Being a semi-numerical code, 
\textsc{21cmSPACE} takes a few hours to evaluate the full cosmic history and the corresponding 21-cm signal, which is too slow to be directly used in the likelihood calculations. To bypass this hurdle, we create emulators of the four observable outputs trained on full runs of \textsc{21cmSPACE}, see subsection~\ref{ssec:emu}. We bring all the above ingredients together in subsection~\ref{sec:inference} where we describe our Bayesian analysis methodology for extracting astrophysical constraints from the four data sets using a joint likelihood.

\subsection{Upper limits on the 21-cm power spectrum}\label{sec:h1c_idr2}

The first observational data we use to constrain properties of the early Universe are the upper limits on the 21-cm power spectrum \citep{HERA_IDR3} from the drift scan radio interferometer, HERA \citep{HERA}. The HERA Phase 1 system used the cross-dipole feeds and the correlator from PAPER \citep{paper_paper, HERA}, allowing measurements from $100-200$ MHz. While data was reduced across the entire frequency range, only two bands were relatively free from radio frequency interference (RFI). These two spectral windows consist of Band 1 from $117.19-133.11$ MHz ($z_\textrm{Band 1}\approx10.35$), and Band 2 from $152.25-167.97$ MHz ($z_\textrm{Band 2}\approx7.87$). 
The HERA Phase 1 measurements were made with 94 nights of observations using 35-41 antennae. We use data from Band 1 in Field D ($6.25-9.25$ hours LST) and from Band 2 in Field C ($4.0-6.25$ hours LST) as they currently provide the best upper limits on the 21-cm power spectrum at \mbox{$\Delta_{21,\textrm{Band 2}}^{2}(k=0.34\ h\mathrm{Mpc}^{-1})\leq 457\ \mathrm{mK}^2$} and \mbox{$\Delta_{21,\textrm{Band 1}}^2(k=0.36\ h\mathrm{Mpc}^{-1})\leq 3496\ \mathrm{mK}^2$}. We decimate the data (illustrated on Figure \ref{fig:data_hera}) using every other $k$-bin to ensure neighbouring data points are uncorrelated in our analysis. The \citet{HERA_IDR3} limits use the same foreground avoidance approach \citep{Kerrigan_2018_foreground_avoidance,Morales_2019_foreground} as the earlier set of HERA Phase 1 limits presented in \citet{HERA_IDR2a_limits}. However due to the longer integration time (more nights of observation), the \citet{HERA_IDR3} limits are 2.6 and 2.1 times deeper relative to the previous best limits at \mbox{$\Delta_{21,\textrm{Band 2}}^2(k=0.192\ h\mathrm{Mpc}^{-1})\leq 946\ \mathrm{mK}^2$} and \mbox{$\Delta_{21,\textrm{Band 1}}^2(k=0.256\ h\mathrm{Mpc}^{-1}) \leq 9166\ \mathrm{mK}^2$}
reported in \citet{HERA_IDR2a_limits}.

\subsection{Global 21-cm signal}~\label{ssec:global}
\begin{figure*}
  \centering
  \begin{subfigure}{0.47\textwidth}
  \centering
    \includegraphics[width=\linewidth]{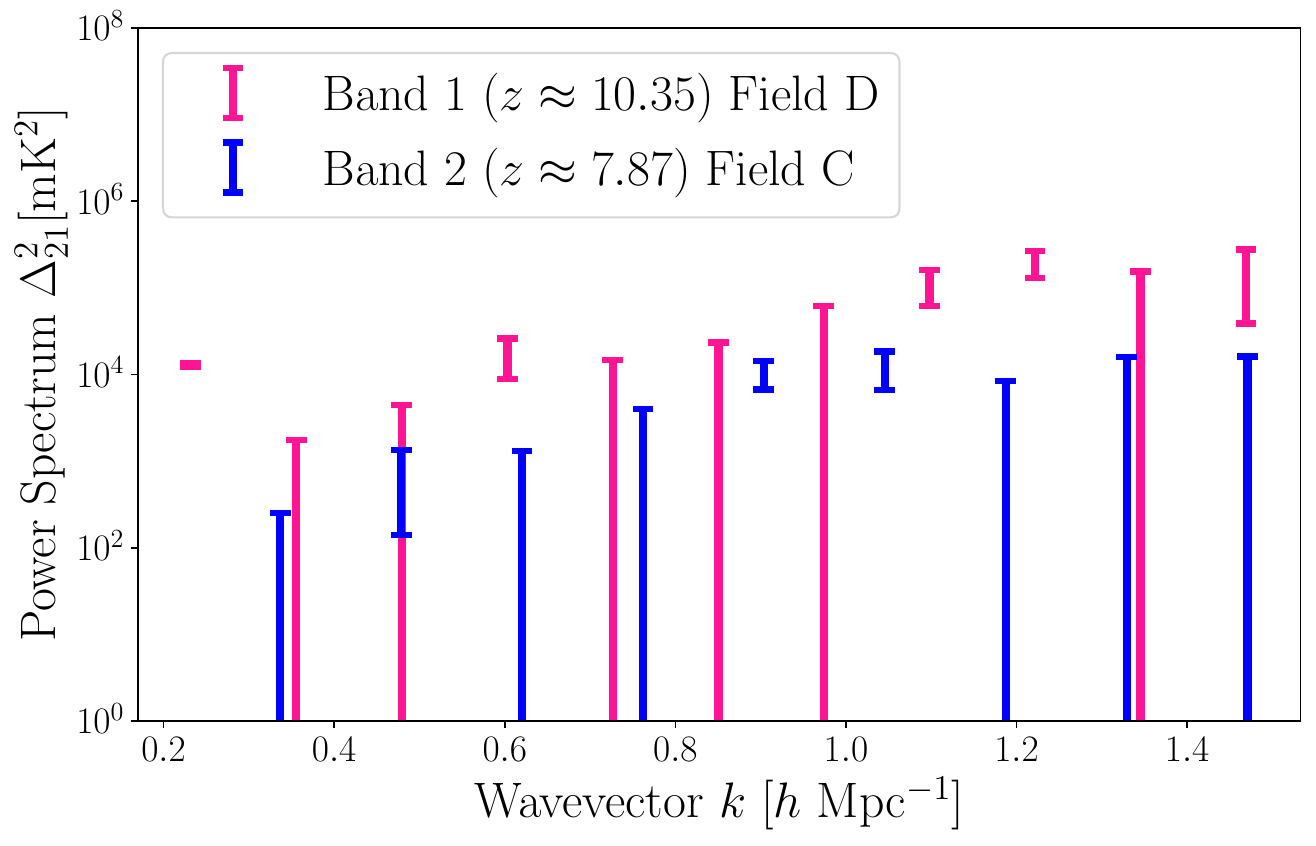}
    \caption{}
    \label{fig:data_hera}
  \end{subfigure}
  \begin{subfigure}{0.49\textwidth}
  \centering
    \includegraphics[width=\linewidth]{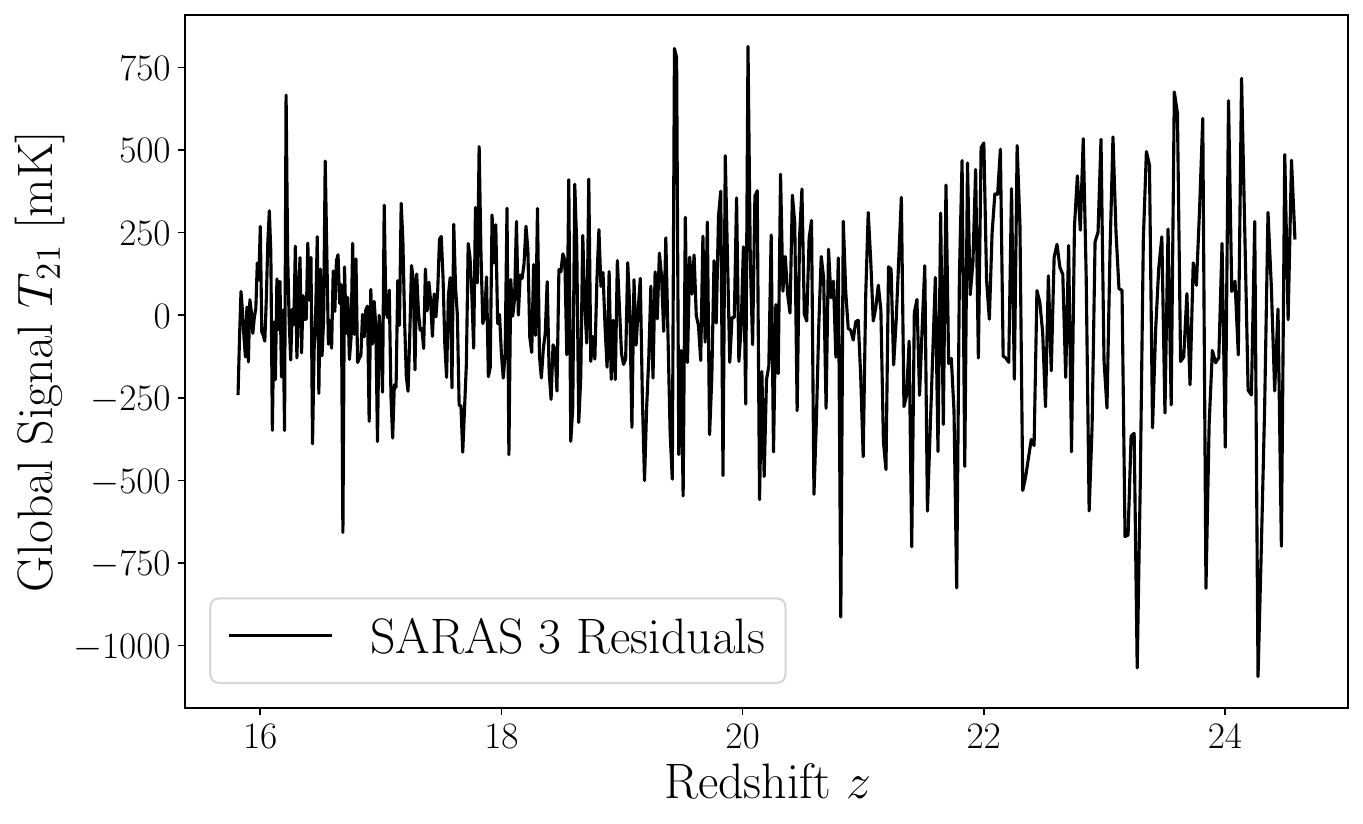}
    \caption{}
    \label{fig:data_saras3}
  \end{subfigure}
  \begin{subfigure}{0.48\textwidth}
  \centering
  \raisebox{1.2cm}{
    \begin{tabular}[b]{ll}
    \hline
    \multirow{2}{*}{Data set} & Integrated X-ray background \\ & $S\ [\mathrm{erg\ cm^{-2}\ s^{-1}\ deg^{-2}}]$ \\
    \hline
    $1-2\ \mathrm{keV}$ & \multirow{2}{*}{$(1.04 \pm 0.14) \times 10^{-12}$} \\
    \citet{hickox2006} &  \\
    $2-8\ \mathrm{keV}$ & \multirow{2}{*}{$(3.4 \pm 1.7) \times 10^{-12}$} \\
    \citet{hickox2006} &  \\
    $8-24\ \mathrm{keV}$ & \multirow{2}{*}{$(1.832 \pm 0.042) \times 10^{-11}$} \\
    \citet{harrison2016} &  \\
    $20-50\ \mathrm{keV}$ & \multirow{2}{*}{$(1.998 \pm 0.083) \times 10^{-11}$} \\
    \citet{harrison2016} & \\
    \hline
    \end{tabular}
    }
    \caption{}
    \label{tab:data_chandra}
  \end{subfigure}
  \begin{subfigure}{0.48\textwidth}
  \centering
    \includegraphics[width=\linewidth]{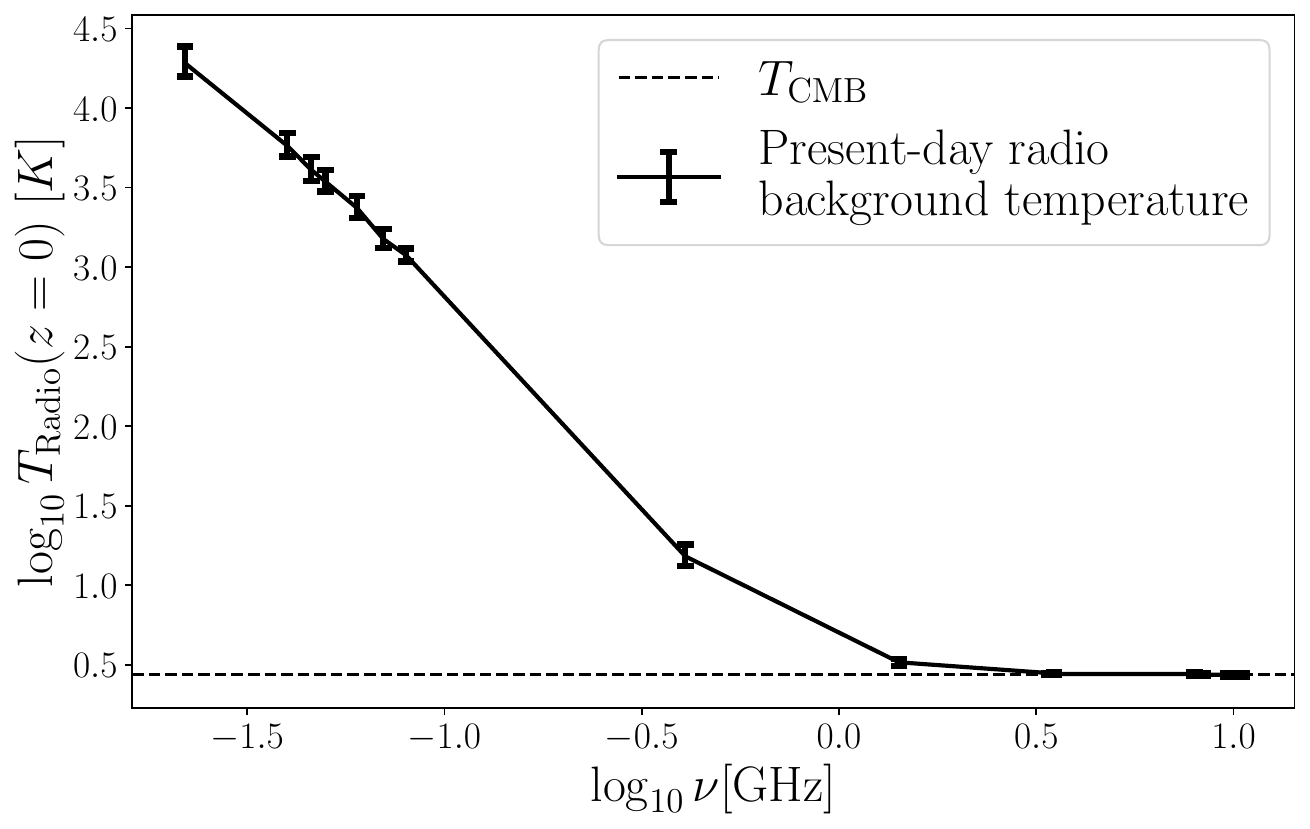}
    \caption{}
    \label{fig:data_lwa}
  \end{subfigure}
  \caption{Observational data used in this study to constrain 21-cm model parameters. \subref{fig:data_hera}) The decimated $1\sigma$ upper limits on the 21-cm power spectrum from HERA in Band 1 ($z\approx10.35$) in pink and Band 2 ($z\approx7.87$) in blue \citep{HERA_IDR3}. We decimate the data using only every other $k$ bin data point to ensure neighbouring data points are uncorrelated in our analysis. 
  \subref{fig:data_saras3}) The residuals after subtracting the best-fit foreground model from the SARAS~3 data \citep{saras3_measurement}.  
  \subref{tab:data_chandra}) Measurements of the unresolved Cosmic X-ray Background in $1-2$ keV and $2-8$ keV (\textit{Chandra}), and $8-24$ keV and $20-50$ keV bands (\textit{HEAO}, \textit{BeppoSAX}, \textit{Swift}).
  \subref{fig:data_lwa}) The measurements of the present-day radio background temperature \citep[solid line with errorbars,][]{tradio_data} and the temperature of the CMB (dashed line).
  }
\end{figure*}

Additionally, we use 15 hours of $55 - 85$ MHz ($z_\textrm{SARAS 3}\sim 15 - 25$) measurements of the global sky temperature \citep{saras3_measurement} from the third generation of the Shaped Antenna measurement of the background RAdio Spectrum \citep[SARAS~3,][]{saras1,saras2,saras3} experiment which is attempting to measure the global 21-cm signal. As the data has been reduced and corrected for environmental and intrinsic antennae and receiver effects, the measured sky temperature is expected to be the sum of the global cosmological 21-cm signal and the foreground temperature from galactic and extragalactic sources along with any residual systematics. On Figure~\ref{fig:data_saras3} we show the residuals of the best-fit foreground subtracted SARAS 3 data. Similar to \citet{saras3_measurement} we model the foreground temperature by a 6th order $\log$-$\log$ polynomial. To fit the foreground alongside the global signal, the polynomial coefficients are sampled from uniform distributions, $a_i \in [-10, 10]$, and the Gaussian noise in the data is sampled from a log-uniform distribution, $\sigma_\mathrm{noise} \in [0.01, 0.5]\ \mathrm{K}$, during the nested sampling.

\subsection{Cosmic X-ray Background}~\label{ssec:xrb}
We also include measurements of the integrated X-ray background (CXB) in four different bands shown in Figure~\ref{tab:data_chandra}. 
As current CXB measurements contain unresolved sources across different redshifts, diffuse emission, and potentially unknown systematics, we treat the following observations as upper limits on the integrated X-ray background. 
The observational data is obtained from \citet{hickox2006} and \citet{harrison2016}. 
Using the \textit{Chandra X-ray Observatory} \citet{hickox2006} measured the unresolved CXB flux as \mbox{$S = (1.04 \pm 0.14) \times 10^{-12}\ \mathrm{erg\ cm^{-2}\ s^{-1}\ deg^{-2}}$} in the $1-2$ keV band and \mbox{$S = (3.4 \pm 1.7) \times 10^{-12}\ \mathrm{erg\ cm^{-2}\ s^{-1}\ deg^{-2}}$} in the $2-8$ keV band. The unresolved CXB intensity in the $1-2$ keV and the $2-8$ keV bands were higher than expected, potentially due to Galactic Local Bubble emission \citep{snowden_bubbles} or a truly diffuse component. 
In higher energy bands, \citet{harrison2016} collated CXB measurements from \citet{gruber_heao} using the \textit{HEAO-1} A2 and A4 instruments, \citet{marshall_heao} using the \textit{HEAO-1} A2 instrument, \citet{frontera_bat} using the \textit{BeppoSAX} and \textit{HXMT} instruments, and \citet{ajello2008} using the \textit{Swift} BAT instrument, among others. For the \textit{BeppoSAX} and \textit{Swift} BAT measurements, which did not cover $8-24$ keV, the CXB intensities were extrapolated using the conversion from Equation 5 in \citet{ajello2008}. To constrain our parameter inference we use the values from \citet{harrison2016} in the $8-24$ and $20-50$ keV bands to determine average upper limit values of $S(8-24\ \mathrm{keV}) \leq (1.832 \pm 0.042) \times 10^{-11}\ \mathrm{erg\ cm^{-2}\ s^{-1}\ deg^{-2}}$ and $S(20-50\ \mathrm{keV}) \leq (2.0 \pm 0.083) \times 10^{-11}\ \mathrm{erg\ cm^{-2}\ s^{-1}\ deg^{-2}}$. 

Given the dependence on the X-ray efficiency, $f_X$, in equation~\eqref{eq:xraybackground} we expect the X-ray Background data to primarily constrain this parameter. However, it is also likely to constrain the star formation efficiency parameters due to the dependence on the star formation rate density. 
While the astrophysical parameter constraints from the X-ray background data are expected to be dominated by the lowest limits from \citet{hickox2006} \textit{Chandra} measurements, the data set also includes measurements from other observatories in other energy bands. Therefore, we will refer to the parameter constraints from the CXB data collectively as constraints from the \textit{X-ray Background} data throughout this study.

\subsection{Excess Radio Background}~\label{ssec:erb}
An excess radio background above the CMB has been reported by the ARCADE2 \citep{arcade2,arcade2_seiffert} and the LWA1 \citep{lwa1, tradio_data} experiment. 
This excess radio background could be the result of unresolved contributions from high redshift radio galaxies \citep{reis2020} although it is not certain that the measured excess radio background has extragalactic origins \citep{subrahmanyan_excess}. We use observational data of the present-day excess radio background temperature from table 2 of \citet{tradio_data} (shown on Figure~\ref{fig:data_lwa}) to constrain our astrophysical parameters. 
As the radio background likely contains low-redshift unresolved sources, we treat these observations as upper limits on the high-redshift contribution to the present-day excess radio background temperature in our parameter inference. 
From the radio background data we expect direct constraints on the radio efficiency parameter, but also the star formation efficiency, given their relationship in equation~\eqref{eq:trad_equation}. 
While the data consists mostly of measurements from the LWA1 Low Frequency Sky Survey \citep{lwa1} below $0.1$ GHz and ARCADE2 \citep{arcade2} from $3-11$ GHz, individual data points from other surveys are also included at $0.022$ GHz \citep{roger_radio}, $0.046$ GHz \citep{alvarez_radio,maeda_radio}, $0.408$ GHz \citep{haslam_radio, remazeilles_radio}, and $1.419$ GHz \citep{reich_radio1,reich_radio2,reich_radio3}. 
For the remainder of this paper, we will refer to the constraints from the measurements of the radio background temperature as constraints from the \textit{Radio Background} data.
\subsection{Training emulators}~\label{ssec:emu}
In order to infer parameter constraints we need to explore the parameter space by varying the astrophysical parameters and compare the simulated observables to the observational data. This requires a large number of simulations (e.g. $\mathcal{O}(2\times10^7)$ likelihood calls for the joint fit), which is not feasible in a reasonable timespan with \textsc{21cmSPACE}. Therefore we train regression multi-layer perceptron neural networks on the outputs from a set of 109,525 simulations (approximately 1~million CPU core hours) with parameter values randomly sampled from within the prior ranges listed in Section \ref{sec:simulations}. This allows us to get predictions of the expected observables in fractions of a second. Specifically we train emulators on the 21-cm power spectrum, the present-day radio background temperature, the X-ray background, and the 21-cm global signal. The architecture of the neural networks is based on, or directly uses, \texttt{globalemu} \citep{globalemu}. Throughout this section we will cover the architecture and accuracy of the trained neural networks used for parameter inference, and we summarise the relevant emulator settings in table~\ref{tab:emu_settings}. 

\subsubsection*{Global 21-cm signal emulator}

The global 21-cm signal emulator was trained using \texttt{globalemu} which is built on \texttt{TensorFlow} \citep{tensorflow2015-whitepaper} and \texttt{Keras} \citep{chollet2015keras}. The advantage of using \texttt{globalemu} is the thorough pre-processing of the input global signal, and the fact that the redshift is taken as an input ensuring continuous global 21-cm signal outputs. The pre-processing involves subtracting an Astrophysics Free Baseline model to simplify the relationship between parameters and signal, and down-scaling and standardising the input by dividing with the standard deviation of the signal. This decreases the complexity of the problem, making it easier for the emulator to learn the relationship between the input parameters and the global signal, and ultimately improves the accuracy. \texttt{globalemu} also allows the user to resample the global signal around the turning points, where the variation in the signal is larger. However, we disable this setting as we manually resample the input data by interpolating to all the SARAS~3 frequencies (redshifts) and at every integer redshift outside the SARAS~3 band. While the full redshift range of the emulator is $z=6-28$, the higher interpolation frequency in the SARAS 3 band helps the emulator learn the signal in the range relevant to the data constraints.

The global signal neural network consists of 4 hidden layers (20~nodes each), uses a $\tanh$ activation function, and \texttt{Adam} \citep{adam} to update the network parameters. We train on batches of 769 models, and of the full set of simulations, we use 66\% (34\%) for training (testing). To prevent overfitting we use early-stopping, where upon training completion, the global signal emulator achieves a 95 percentile root-mean-square error of 40 mK and a 68 percentile accuracy of 20\%. 

\subsubsection*{21-cm power spectrum emulator}
Our $\Delta_{21}^2$ power spectrum emulator utilises a very similar approach to the global 21-cm signal emulator. 
While the network architecture resembles that of \texttt{globalemu}, the power spectrum emulator is built using the \texttt{scikit-learn} \citep{sklearn} multi-layer perceptron regression neural network. The power spectrum emulator takes the redshift, the magnitude of the wavevector, and the simulation parameters ($2+9$ parameters in total) as inputs, and outputs a $\Delta_{21}^{2}$ power spectrum value as in \citet{HERA_IDR2b_theory} and \citet{HERA_IDR3} (Section 8.3 and Section 7.5 respectively). 
The input data is pre-processed before training to improve the emulator performance. The pre-processing simplifies the relationship between the input and output, which again helps increase the emulator accuracy. First the simulation set was split such that 80\% was used for training and 20\% was used for testing to assess the quality of the trained emulator. 
The training data was then resampled and interpolated to a finer $zk$-grid by drawing $N_{\mathrm{over}}=500$ pairs of $z$-values and $k$-values from uniform distributions over the ranges $z=7-26$ and $k \approx 0.1 - 1.5\ h\mathrm{Mpc^{-1}}$ for each simulation, as this was the $k$-range output by the simulation code.  
To avoid the emulator putting too much weight on learning low power models, at the expense of accuracy with models closer to the relevant HERA upper limits, the input was truncated such that data points below $1\ \mathrm{mK^2}$ were set to $1\ \mathrm{mK^2}$. Then the input was log transformed, to ensure better performance across the entire dynamic range. This log transformation was then reversed to get the power spectrum output from the trained network. The training was conducted using the \texttt{MLPRegressor} with 4 hidden layers with 100 nodes each, a \texttt{ReLU} activation function \citep{relu} between layers, the \texttt{Adam} parameter optimizer, a constant learning rate of 0.001, and a batch size of 200 models to pass through the network during training. This resulted in a 95 percentile accuracy of $\sim 20\%$ in the HERA upper limit bands, which is similar to the relative emulator error of 20\% used in \citet{HERA_IDR2a_limits} (Section 8.3).

\subsubsection*{X-ray background \& radio background temperature emulator}
We trained the emulator on the X-ray background simulation data, taking the same approach as with the power spectrum emulator. The X-ray background simulation data set was split in to an 80\% training batch and a 20\% test batch. The training batch was then resampled by drawing $N_{\mathrm{over}}=400$ energies to linearly interpolate X-ray background data points in $\log-\log$ space. A regression network with 4 hidden layers (50 nodes each), \texttt{ReLU} activation, and \texttt{Adam} parameter optimization was then trained on the simulation parameters and $\log$ energies (9+1 parameters). The X-ray background emulator achieved an accuracy of $5\%$ at the 99 percentile level. 

The present-day radio background temperature is calculated in the post-processing by integrating over the contribution of all galaxies in the past lightcone \citep[equation~\eqref{eq:trad_equation},][]{reis2020}. 
We train an emulator on the derived radio background temperature following a similar procedure to the X-ray background emulator, using an 80\% (20\%) training (test) split, drawing $N_{\mathrm{over}}=200$ frequencies for resampling and interpolation, using 4 hidden layers (50 nodes each) with \texttt{ReLU} activations, and \texttt{Adam} optimization. The emulator was trained on the simulation parameters and the radio background frequencies (9+1 parameters). This resulted in a 95 percentile accuracy of $5\%$.

\begin{table*}
\centering
\caption{
Neural network settings used to train emulators of the 21-cm power spectrum, X-ray background, present-day radio background temperature, and 21-cm global signal. The emulators were developed using different frameworks and network architectures. The batch size, activation function, parameter optimisation scheme, train (test) split percentage, learning rate, and accuracy are provided for each emulator.
}
\label{tab:emu_settings}
\begin{tabular}{llllllllll}
\hline
Emulator & Framework & Network architecture & Batch size & Activation & Optimizer & Train (Test) \% & Learning rate & Accuracy\\
\hline
Power Spectrum & \texttt{scikit-learn} & 11-100-100-100-100-1 & 200 & ReLU & Adam & 80 (20) &  0.001 & $20\%$\\
CXB & \texttt{scikit-learn} & 10-50-50-50-50-1 & 200 & ReLU & Adam & 80 (20) &  0.001 & $5\%$\\
$T_\mathrm{r}$ & \texttt{scikit-learn} & 10-50-50-50-50-1 & 200 & ReLU & Adam & 80 (20) &  0.001 & $5\%$\\
Global Signal & \texttt{globalemu} & 10-20-20-20-20-1 & 769 & $\tanh$ & Adam & 66 (34) & 0.001 & $20\%$ \\
\hline
\end{tabular}
\end{table*}

\subsection{Bayesian inference} \label{sec:inference}
For our parameter inference we take a Bayesian approach calculating the posterior probability of the model parameters, $\theta$, given the observed data, $\mathcal{D}$,
\begin{equation}
    P(\theta|\mathcal{D}) = \frac{\mathcal{L}(\theta)\pi(\theta)}{\mathcal{Z}}.
\end{equation}
Here the Bayes theorem is expressed in terms of the likelihood, $\mathcal{L}(\theta)=P(\mathcal{D}|\theta)$, the prior belief of the probability of the parameters $\pi(\theta)=P(\theta)$, and the Bayesian evidence $\mathcal{Z} = P(\mathcal{D})$. The model parameter vector may contain parameters that we consider nuisance parameters, as we are just interested in comparing the posterior probabilities of the astrophysical parameters constrained by different experiments. The parameters can thus be split into astrophysical parameters, $\theta_a$, and nuisance parameters $\theta_n$, e.g. $\theta = \{\theta_{a}, \theta_\mathrm{n}\}$. We can then marginalise over the nuisance parameters to get the posterior probability of the astrophysical parameters alone. If the prior is uniform the problem further simplifies as the posterior probability becomes proportional to the likelihood, 
\begin{equation}
    P(\theta_a | \mathcal{D}) \propto \mathcal{L}(\theta_a) .
\end{equation}
Bayesian nested sampling \citep{ns_skilling, ns_ashton} iteratively evolves live points to increasingly higher likelihoods, allowing us to determine the Bayesian evidence while sampling the posterior probability distribution function to produce astrophysical parameter constraints. To conduct nested sampling, we use the slice-sampling algorithm implemented in \texttt{PolyChord} \citep{polychord1,polychord2}. 

For the upper limits from HERA, we use the likelihood function marginalised over unknown positive systematics as in \citet{HERA_IDR2b_theory}, 
\begin{equation} \label{eq:likelihood}
    \mathcal{L}(\theta) = \prod_{i}^{N_\mathrm{d}} \frac{1}{2} \left(1+\mathrm{erf}\left[\frac{d_i - m_i (\theta)}{\sqrt{2(\sigma_{d_i}^2 + \sigma_{m_i}^2)}}\right] \right).
\end{equation}
Here $N_\mathrm{d}$ is the number of data points, $d_i$ is the observational data, $m_i (\theta)$ is the corresponding modelled power spectrum value generated from the particular set of simulation parameters, $\theta$, and $\sigma_{d_i}$ and $\sigma_{m_i}$ are the corresponding data and model errors. 
A likelihood of the same form is used for the CXB and the present-day radio background temperature as we also treat these as upper limits.
The error in the theoretical power spectrum model is set to $20\%$, and the model error for the CXB and present-day radio background temperature is set to $5\%$ according to the emulator precision. 
Equation \ref{eq:likelihood} assumes the error on each data point is uncorrelated. However, as the HERA $k$ window functions overlap we take the same approach as \citet{HERA_IDR2b_theory} to ensure uncorrelated data points and decimate the data, using only every other $k$ bin. Specifically, we pick the lowest data point (tightest constraint) and include every other data point below and above that $k$ value for each band. 

Lastly, for the SARAS~3 data we take the same approach as \citet{saras3_measurement} and \citet{bevins_saras3,bevins_joint_paper} and model the foreground temperature by a 6th order $\log$-$\log$ polynomial,
\begin{equation} \label{eq:T_fg}
    \log_{10} \left(T_\mathrm{fg}\right) = \sum_{i=0}^{i=6} a_i \left( \mathcal{R}(\log_{10}\nu) \right)^{i}.
\end{equation}
Where $\mathcal{R}(\log_{10}\nu)$ linearly scales the log frequency to be from $-1$ to $1$ and $a_i$ are the foreground polynomial coefficients. 
We fit the foreground polynomial coefficients and the signal noise (which we consider nuisance parameters in this study) alongside the simulation parameters. Following \citet{bevins_saras3, bevins_joint_paper} we adopt a Gaussian likelihood of the form
\begin{equation}
\begin{split}
    \log \left(\mathcal{L}_\mathrm{SARAS\ 3}(\theta)\right) = \sum_{i}^{N_\mathrm{d}} 
    \left(
    -\frac{1}{2} \log\left(2\pi\left(\sigma_\mathrm{noise}^2 + \sigma_\mathrm{signal}^2\right)\right)
    \right. \\
    \left.
    -\frac{1}{2} 
    \frac{(T_\mathrm{SARAS\ 3,i} - T_\mathrm{fg,i} - T_\mathrm{signal,i})^2}{(\sigma_\mathrm{noise}^2+\sigma_\mathrm{signal}^2)}
    \right).
\end{split}
\end{equation}
Where $T_\mathrm{signal,i}$ is the model signal at frequency $\nu_i$, $T_\mathrm{SARAS\ 3,i}$ is the global sky temperature measured by SARAS~3, $T_\mathrm{fg,i}$ is the foreground temperature given by Equation \ref{eq:T_fg}, and $\sigma_\mathrm{noise}$ is the noise parameter. Unlike \citet{bevins_saras3,bevins_joint_paper} we include the uncertainty on the modelled signal, $\sigma_\mathrm{signal}$, which arises from the emulator imprecision. This modelling error on the global signal is set to $20\%$ due to the aforementioned emulator accuracy.

In nested sampling runs with multiple observational constraints simultaneously imposed, we treat each observation as independent such that the total likelihood of a sample is the product of the individual likelihood contributions from each constraint. When all observational data is included, the total likelihood is hence given by 
\begin{align}
\begin{split}
\mathcal{L}_{\mathrm{total}} &= \mathcal{L}_\mathrm{HERA} \times \mathcal{L}_\textrm{X-ray Background} \times \mathcal{L}_\textrm{Radio Background} \\ &\times \mathcal{L}_\mathrm{SARAS\ 3}. 
\end{split}
\end{align}
\section{Results} \label{sec:results}

Having established our 21-cm signal model, the architecture of our emulators, analysis methodology, and the observational data used, we now present the results from our analysis. 
We use \texttt{anesthetic} \citep{anesthetic} and \texttt{fgivenx} \citep{fgivenx} to read the chains, plot prior and posterior samples to illustrate the parameter constraints, and plot functional posteriors.
Five separate nested sampling runs were conducted - one for each observational data set, and another with the constraints jointly imposed.

\begin{figure*}
\includegraphics[width=\textwidth]{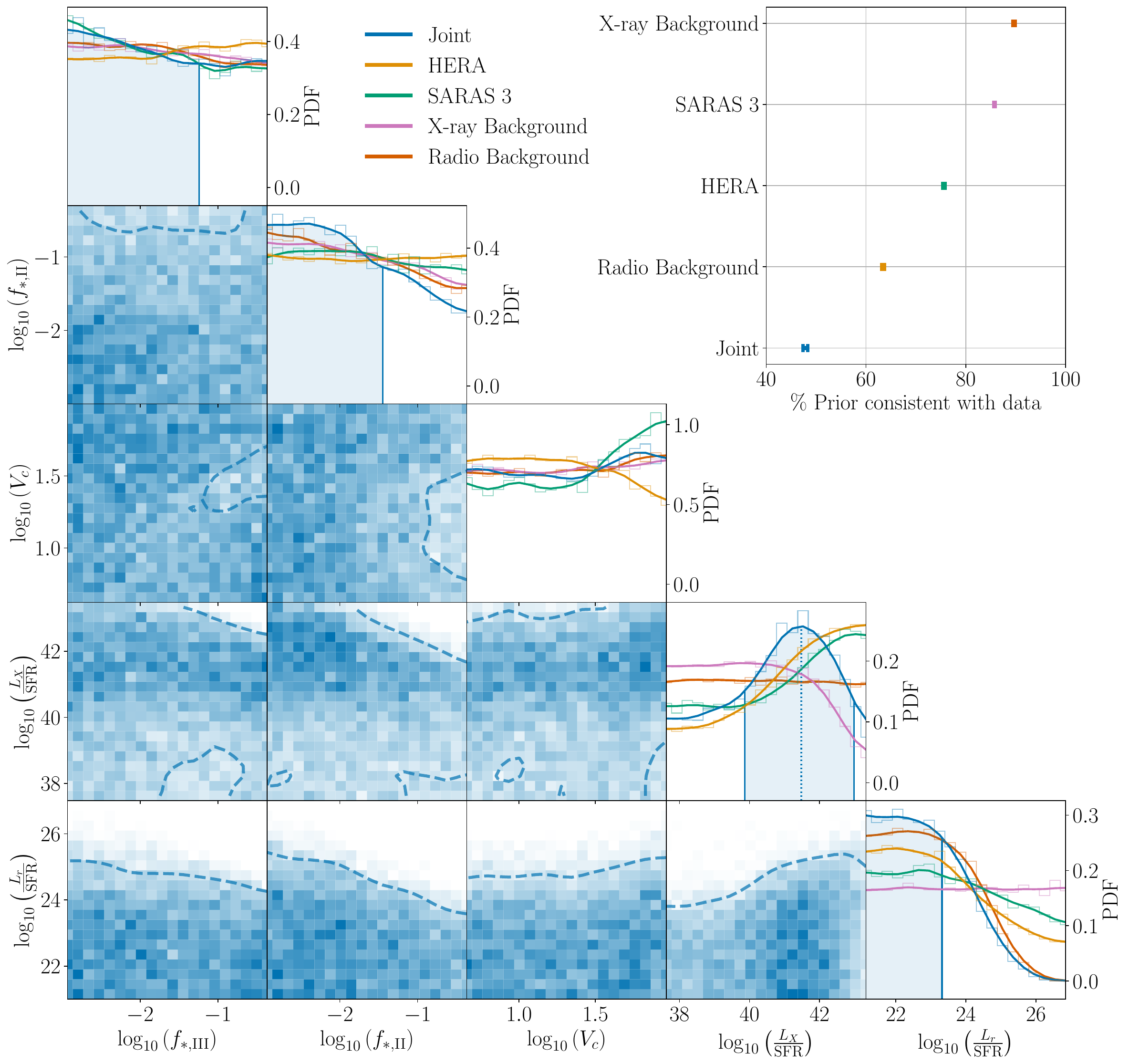}
\caption{Constraints obtained from the Bayesian analyses for the individual experiments: HERA \citep[orange, ][]{HERA_IDR3}, SARAS 3 \citep[green, ][]{saras3_measurement}, the X-ray Background \citep[pink, ][]{hickox2006,harrison2016}, and the Radio Background \citep[red, ][]{tradio_data}. The joint constraints are shown in blue and include all the data sets in the likelihood calculation. The diagonal section of the triangle plot shows the 1D marginal posterior PDFs for each individual run alongside 68 percentile confidence intervals on parameters constrained by the joint fit (vertical blue lines). Our analysis provides one of the first hints of a constraint on early Pop III star hosting galaxies, with the 1D marginal posterior on the Pop III star formation efficiency showing a 68 percentile preference for $f_\mathrm{\ast,III}\lesssim 5.7\%$. 
In addition, we find a posterior peak in the 1D marginal posterior PDF on the logarithm of X-ray luminosity per star formation rate at $\approx 41.48$ and a 68 percentile confidence interval around the weighted average, $\approx 40.77_{-0.90}^{+2.21}$. 
The 2D marginal posterior PDFs (bottom left half of the triangle plot) show the constraints from the joint fit with dashed contours indicating the 95\% confidence level (the corresponding 2D PDFs for each individual experiment can be found in Figure  \ref{fig:frad_joint_individuals}). Here we see degeneracies: e.g. high values of $f_\mathrm{\ast,III}$ are ruled out together with high $L_X/\mathrm{SFR}$ and high $L_\mathrm{r}/\mathrm{SFR}$. Additionally, in agreement with previous works, we see that the combination of low X-ray and high radio efficiencies is ruled out by the 21-cm data \citep{HERA_IDR2b_theory, bevins_joint_paper}. The disfavoured region extends to higher X-ray and radio efficiencies due to the synergistic addition of the X-ray and Radio Background data. In the top right panel, we show the percentage of the prior consistent with the data for each of the individual, and the joint, analysis. The advantage of the joint analysis is evident as the joint fit compresses the prior space the most to $47.8^{+0.5}_{-0.5}\%$ of the prior volume. 
\label{fig:frad_joint_all}}
\end{figure*}

Figure \ref{fig:frad_joint_all} shows the 1D and 2D marginal posterior probability density functions (PDFs) of the astrophysical parameters for the joint analysis (1D PDFs are also shown for each experiment individually, while the corresponding 2D PDFs can be found in the appendix, Figure  \ref{fig:frad_joint_individuals}). The triangle plot illustrates the constraining power of each observational data set, as well as the joint constraints (blue lines), across the simulated parameter ranges. Upper and lower 68 (95) percentile confidence regions on the 1D marginal parameter posteriors are summarised in table \ref{tab:quantiles}. We also show the percentage of the explored astrophysical prior space consistent with each data set in the top right corner of Figure \ref{fig:frad_joint_all}.  

\begin{table*}
\caption{Upper and lower 68 (95) percentile confidence regions on 1D marginal parameter posteriors. While this table presents values that can be used to infer constraints on the astrophysical parameters, all of these should not be interpreted as upper or lower limits as the majority of these values are still prior dominated (as illustrated by their proximity to the prior value). The values highlighted in bold appear to be likelihood dominated as they differ more from the prior values. These highlighted values can be interpreted as a 68 (95) percentile disfavouring of part of the astrophysical parameter space.}
\label{tab:quantiles}
\begin{tabular}{l c c cc cc c}
 & $f_\mathrm{\ast,III}\ [\%]$ & $f_\mathrm{\ast,II}\ [\%]$ 
 & \multicolumn{2}{c}{$V_c\ [\mathrm{km\ s^{-1}}]$} 
 & \multicolumn{2}{c}{$\log_{10} L_X/\mathrm{SFR}$} & $\log_{10} L_\mathrm{r}/\mathrm{SFR}$ \\
 Prior range & 
 \multicolumn{1}{c}{$[0.1, 50.0]$} & 
 \multicolumn{1}{c}{$[0.1, 50.0]$} & 
 \multicolumn{2}{c}{$[4.2, 100.0]$} & 
 \multicolumn{2}{c}{$[37.48, 43.48]$} & 
 \multicolumn{1}{c}{$[21, 27]$} \\
68\% (95\%) Confidence & Upper & Upper & Lower & Upper & Lower & Upper & Upper \\ \hline
{Prior} & $\leq6.8 (36.6)$ & $\leq6.8 (36.6)$ & $\geq11.58 (4.92)$ & $\leq36.26 (85.34)$ & $\geq39.40 (37.78)$ & $\leq41.56 (43.18)$ & $\leq25.08 (26.70)$  \\ 
{Joint} & $\leq\mathbf{5.7} (35.6)$ & $\leq\mathbf{3.5 (28.7)}$ & $\geq11.99 (4.94)$ & $\leq40.27 (86.45)$ & $\geq40.14 (37.94)$ & $\leq41.68 (42.95)$ & $\leq\mathbf{23.32 (24.77)}$  \\ 
{HERA} & $\leq7.5 (37.1)$ & $\leq7.0 (37.0)$ & $\geq10.76 (4.88)$ & $\leq\mathbf{31.33} (79.75)$ & $\geq\mathbf{40.49} (38.06)$ & $\leq42.23 (43.29)$ & $\leq\mathbf{23.99 (26.30)}$  \\ 
{X-ray Background} & $\leq6.2 (35.6)$ & $\leq5.2 (33.8)$ & $\geq11.78 (4.96)$ & $\leq37.52 (86.01)$ & $\geq39.14 (37.73)$ & $\leq\mathbf{40.98 (42.64)}$ & $\leq25.07 (26.69)$  \\ 
{Radio Background} & $\leq5.8 (35.5)$ & $\leq4.6 (33.5)$ & $\geq12.06 (4.96)$ & $\leq38.96 (86.76)$ & $\geq39.38 (37.78)$ & $\leq41.53 (43.17)$ & $\leq\mathbf{23.56 (24.97)}$  \\ 
{SARAS3} & $\leq\mathbf{5.2} (35.4)$ & $\leq5.8 (34.9)$ & $\geq\mathbf{13.71} (5.01)$ & $\leq46.93 (89.46)$ & $\geq\mathbf{40.03} (37.86)$ & $\leq42.15 (43.27)$ & $\leq\mathbf{24.57} (26.53)$  \\  
\end{tabular}
\end{table*}

\subsection{Consistency of the explored prior volume with data}
As in \citet{bevins_joint_paper, thomas2024_cosmic_strings}, we quantify how much the astrophysical prior contracts to the posterior volume due to each individual data constraint as well as the joint constraint. In order to compare nested sampling runs with different free parameters we first need to marginalise over nuisance parameters, such as the foreground, noise, and discrete signal parameters (which have a secondary impact on the signals). This leaves us with the astrophysical parameters of interest, $\theta = \{ f_\mathrm{\ast,II}, f_\mathrm{\ast,III}, V_c, f_X, \tau, f_r \}$. We use \texttt{margarine} \citep{margarine2,margarine1} to learn the marginal posterior PDFs of the astrophysical parameters by training masked autoregressive flows. With the trained normalizing flows we can sample marginal posteriors - that are not conditioned on nuisance parameters - to calculate marginal statistics like the Kullback-Liebler (KL) divergence. The KL divergence can then be used to estimate the percentage of the prior consistent with the data via, 
\begin{equation}\label{eq:prior_consistency_eq}
    \%\ \mathrm{consistent}= 100 \times \exp(-\mathcal{D}) \approx 100\times \frac{V_\mathrm{post}}{V_\mathrm{prior}} .
\end{equation}
Here $\mathcal{D}$ is the KL divergence, and $V_\mathrm{post}$ and $V_\mathrm{prior}$ are the posterior and prior parameter volume respectively. 
Equation \ref{eq:prior_consistency_eq} allows us to quantify which observational data set rules out most of the prior parameter volume as illustrated on the top right corner plot of Figure \ref{fig:frad_joint_all}. We find that the joint constraints on the model parameters compress the parameter space the most with  
$47.8^{+0.5}_{-0.5}\%$ 
of the prior being consistent with the data. While this constitutes an improvement from the $64.9^{+0.3}_{-0.1}\%$ prior consistency of the joint HERA and SARAS 3 analysis in \citet{bevins_joint_paper}, the underlying models are different and therefore the prior consistencies are not directly comparable. 
Individually, the Radio Background, HERA, SARAS 3, and X-ray Background data show a 
$63.5^{+0.3}_{-0.3}$, $75.5^{+0.4}_{-0.5}$, $85.9^{+0.1}_{-0.2}$, and $89.7^{+0.2}_{-0.3}$ 
prior consistency with the data respectively. The fact that the joint analysis has the tightest constraints on the prior parameter space justifies the synergistic approach adopted in this work. 

\subsection{Constraints from individual observational data sets}
The novelty of this work is in constraining Cosmic Dawn signals with multi-wavelength data. Most importantly, we find that data collected by the SARAS~3 experiment in the $z\sim 15-25$ redshift range are sensitive to the process of star formation in the early Universe preferring low values of $f_\mathrm{\ast,III}$, low  $f_\mathrm{\ast,II}$, and high  $V_c$. The rest of our results are in broad agreement with previous publications \citep[e.g.][]{bevins_joint_paper, HERA_IDR2b_theory, HERA_IDR3}. In the following, we discuss the 1D and 2D marginal posterior PDFs of Figure \ref{fig:frad_joint_all} in more detail.  
For completeness, triangle plots of the full 1D and 2D marginal posterior PDFs for each experiment are shown on Figure \ref{fig:frad_joint_individuals}. 
Table \ref{tab:quantiles} shows 68 (95) percentile confidence regions on simulation parameters, and in the following we will discuss some of the strongest (highlighted) 68 percentile constraints for each experiment.

The 21-cm data depends on several astrophysical processes \citep[e.g.][]{cohen2017, thomas_popIIIimf} and, thus, upper limits from HERA and SARAS 3 residuals allow us to disfavor a complex combination of parameters. 
\begin{itemize}
    \item HERA: Strong power spectra above the HERA limits are obtained for low X-ray efficiencies and high radio efficiencies. Therefore, the upper limits on the 21-cm power spectrum from HERA result in individual (1D) constraints on X-ray and radio efficiencies:
    \begin{equation*}
    \log_{10} \left(L_X/\mathrm{SFR}\ \mathrm{\left[erg\ s^{-1}M_{\odot}^{-1}yr\right]}\right) \geq 40.49\qquad
    (f_X \geq 1.047),
    \end{equation*}
    and 
    \begin{equation*}
    \log_{10} \left(L_\mathrm{r}/\mathrm{SFR}\ \mathrm{\left[W\ Hz^{-1}M_{\odot}^{-1}yr\right]}\right) \leq 23.99\qquad (f_r \leq 100).
    \end{equation*}
In addition, models corresponding to combinations (2D) of such parameters are ruled out, as can be seen on the corresponding 2D PDFs of Figure \ref{fig:frad_joint_individuals}. The 21-cm power spectrum at the redshifts and scales observable by HERA is also sensitive to the timing of Pop II formation which is regulated by the minimal circular velocity of star-forming halos, $V_c$. High $V_c$ corresponds to a high critical halo mass for star formation, which results in late star formation and strong fluctuations within the HERA band. We find that HERA disfavours $V_c \geq 30\mathrm{\left[km\ s^{-1}\right]}$ (corresponding to a halo of mass $\geq 3\times 10^8M_\odot$ at $z=7$). This is also reflected in the disfavoured regions of the 2D marginal posterior PDFs showing that combining the high star formation thresholds with low X-ray efficiencies or high radio efficiencies leads to a high amplitude power spectrum in the HERA bands, which can be ruled out. 
    \item SARAS 3: 
    As we mentioned above, the SARAS 3 data constrain Pop III star formation giving preference to low values of star formation efficiency $f_\mathrm{\ast,III} \leq 5.2\%$ as well as high values of $V_c \geq 14 \mathrm{\left[km\ s^{-1}\right]}$ (corresponding to a halo of mass $\geq 7\times 10^6 M_\odot$ at $z=20$). Since low virial velocities and high Pop III star formation efficiencies lead to rapid early star formation, in models with such stellar properties the 21-cm absorption trough is located within the constrained redshift range $z_\textrm{SARAS3}\sim15-25$. Coupled with low X-ray efficiencies and high radio efficiencies the trough deepens beyond the SARAS 3 residuals and the data are able to rule out such models.
    
    Similarly to HERA, low X-ray efficiencies and high radio efficiencies are constrained by SARAS 3 as such parameters lead to deep absorption troughs that are easier to rule out. We find that the 1D marginal posterior PDFs show preferences for  
    \begin{equation*}
    \log_{10} \left(L_X/\mathrm{SFR}\ \mathrm{\left[erg\ s^{-1}M_{\odot}^{-1}yr\right]}\right) \geq 40.03\qquad (f_X \geq 0.355), 
    \end{equation*}
    and 
    \begin{equation*}
    \log_{10} \left(L_\mathrm{r}/\mathrm{SFR}\ \mathrm{\left[W\ Hz^{-1}M_{\odot}^{-1}yr\right]}\right) \leq 24.57\qquad (f_r \leq 370).
    \end{equation*}
    The 2D marginal posterior PDF (Figure \ref{fig:frad_joint_individuals}) also shows a disfavoured (although to a lesser degree than with HERA) region of low $f_X$ and high $f_r$. 
  \end{itemize}  
    
The dependence of X-ray and Radio Backgrounds on astrophysical parameters is much simpler than that of the 21-cm signal:
\begin{itemize}
\item X-ray Background: As the X-ray Background is directly related to the X-ray efficiency parameter (Equation \ref{eq:xraybackground}), the strongest constraint provided by this data set is an upper limit on $L_X/SFR$:
    \begin{equation*}
    \log_{10} \left(L_X/\mathrm{SFR}\ \mathrm{\left[erg\ s^{-1}M_{\odot}^{-1}yr\right]}\right) \leq 40.98\qquad (f_X \leq 3.16).
    \end{equation*}
    The X-ray Background data set also shows a preference for $f_\mathrm{\ast,II}\leq 5.2\%$ and a very slight preference for $f_\mathrm{\ast,III}\leq 6.2\%$. This is due to high star formation efficiencies leading to larger stellar populations, which naturally contribute more to the X-ray budget, making these models more likely to exceed the upper limits. Consequently, regions of the 2D marginal posterior PDFs corresponding to high X-ray efficiency as well as high Pop II and Pop III star formation efficiencies are ruled out (white) as is seen in Figure \ref{fig:frad_joint_individuals}. 

    \item Radio Background: The Radio Background is directly related to the radio efficiency parameter through its dependence on the radio luminosity (Equation \ref{eq:trad_equation}), so the strongest constraint provided by this data set is an upper limit on 
    \begin{equation*}
     \log_{10} \left(L_\mathrm{r}/\mathrm{SFR}\ \mathrm{\left[W\ Hz^{-1}M_{\odot}^{-1}yr\right]}\right) \leq 23.56 \qquad (f_\mathrm{r} \leq 36).
    \end{equation*}
    This is consistent with the analysis conducted by \citet{reis2020} where they found the Radio Background data to disfavour~\mbox{$f_\mathrm{r} \gtrsim 100\ (L_\mathrm{r}/\mathrm{SFR} \gtrsim 10^{24} \mathrm{\left[W\ Hz^{-1}M_{\odot}^{-1}yr\right]})$}, which is also suggested by our analysis as seen in the 1D marginal radio efficiency PDF in Figure \ref{fig:radiobackground_triangle}. 
    Combined with a slight preference for $f_\mathrm{\ast,III} \leq 5.8\%$ and $f_\mathrm{\ast,II} \leq 5.2\%$, this leads to ruled-out regions of the 2D marginal posterior PDFs corresponding to high radio efficiency and Pop II and Pop III star formation efficiency. The slight preference for low $f_\mathrm{\ast,II}$ and low $f_\mathrm{\ast,III}$ is explained by the fact that efficient Pop II and Pop III star formation leads to a stronger radio background at a fixed value of $f_r$. 
\end{itemize}

While we find some likelihood dominated parameters at the 68~percentile level, the upper and lower limits weaken at the 95 percentile level as more parameters look prior dominated (illustrated by their similarity to the 95 percentile prior values) in table \ref{tab:quantiles}. To improve these astrophysical parameter constraints we perform a joint analysis including all the data sets.  

\subsection{Parameter constraints from the joint observational data sets}\label{sec:results_joint}
\begin{figure}
\includegraphics[width=\columnwidth]{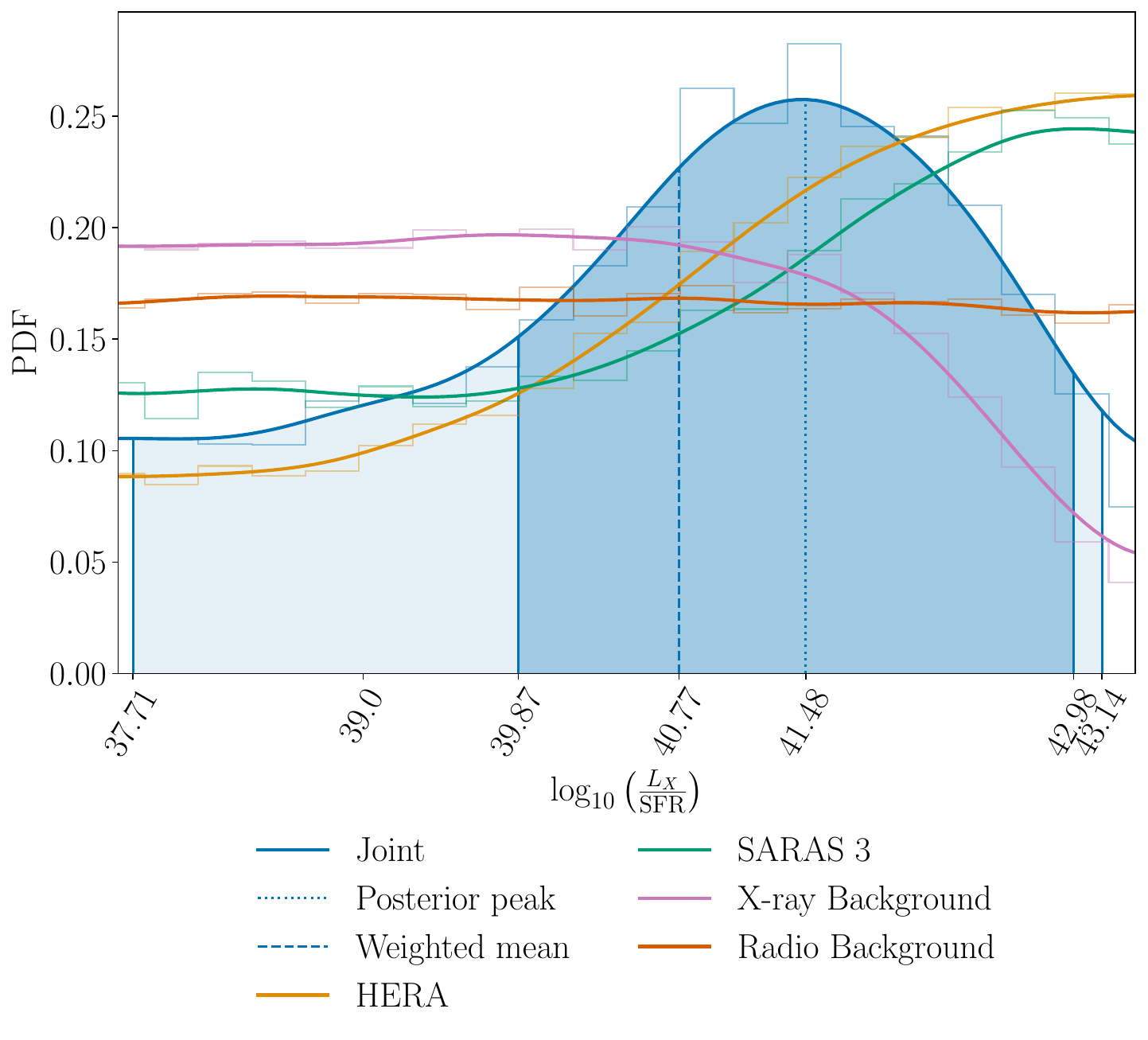}
\caption{X-ray efficiency parameter posterior. The 21-cm data from HERA (orange) and SARAS 3 (green) disfavour low X-ray efficiency models with 68\% lower limits on the $\log$ X-ray luminosity per star formation rate, $\geq 40.49$ for HERA and $\geq 40.03$ for SARAS 3. The X-ray Background data (pink) sets a 68 percentile upper limit $\leq 40.98$, and the Radio Background data (red) leaves the X-ray efficiency unconstrained.
The joint fit (blue) reveals a distinct posterior peak (dotted vertical line) in the $\log$ X-ray luminosity per star formation rate at $\approx 41.48$, with a 68\% (95\%) confidence interval (solid blue vertical lines) around the weighted mean (dashed blue vertical line) of $\log_{10} \left( L_X/\mathrm{SFR}\ \mathrm{\left[erg\ s^{-1}M_{\odot}^{-1}yr\right]} \right) \approx 40.77_{-0.90\ (-3.06)}^{+2.21\ (+2.37)}$. The posterior peak appears due to the X-ray data and 21-cm data constraining opposite ends of the X-ray efficiency prior.
\label{fig:fx_constraints}}
\end{figure}

The combined data constraints provide stronger prior compression compared to the ones derived using each experiment separately. This can be inferred from table \ref{tab:quantiles} as well as the 1D and 2D marginal posterior PDFs of Figure \ref{fig:frad_joint_all}. 
Starting with the 1D marginal posterior PDFs we notice a few interesting features at 68\% confidence. 
\begin{itemize}    
    \item $f_\mathrm{\ast,III}$: We find that in combination the data favour star formation efficiencies of Pop III stars $\lesssim 5.7\%$. This is a slightly weaker upper limit than $\lesssim 5.2\%$ suggested by the SARAS 3 data, and is due to the HERA data showing a very slight preference towards higher star formation efficiencies. The favoured low Pop III star formation efficiencies are consistent with predictions from theory \citep{gurian2023_fstarIII_theory} and numerical simulations \citep{hirano2015_fstarIII_num,jaacks2019_fstarIII_num} that suggest (although with large uncertainties) Pop III star formation efficiencies of the order $f_\mathrm{\ast,III}\sim 0.3\%$ \citep{Klessen_2023}.

    \item $f_\mathrm{\ast,II}$: Here we see a disfavouring of high Pop II star formation efficiencies. While each data set imposes very weak constraints on $f_\mathrm{\ast,II}$, the cumulative preference of all the experiments together results in a somewhat stronger upper limit on  Pop II star formation efficiency $\lesssim 3.5\%$. While it is hard to compare Pop II star formation efficiencies across different models, due to star formation prescription differences and large uncertainties, observations seem to suggest low Pop II star formation efficiencies of the order $f_\mathrm{\ast,II}\sim 1\%$ \citep{behroozi2020_popII_sfe}.

    \item $V_c$: The trends seen in the 1D marginal posterior PDF on $V_c$ from the individual experiments cancel in the joint fit and lead to no joint constraint on $V_c$. The effects of the HERA and SARAS 3 data effectively cancel each other with one experiment disfavouring and the other preferring high values of the virial velocities. This is because the HERA constraints are at relatively low redshift, thus preferring models with low power spectra corresponding to models with low values of $V_c$. In such models stars are created earlier and in smaller and more numerous dark matter halos, resulting in more homogeneous backgrounds. Meanwhile, it is harder for SARAS 3 to reject models with high $V_c$ as in these models the onset of star formation is delayed and the IGM heating by X-ray binaries happens later, resulting in the 21-cm signal trough outside the SARAS 3 band.
    
    \item $L_X/\mathrm{SFR}$: In Figure \ref{fig:fx_constraints} we show an enlarged version of the 1D posterior PDF on the X-ray luminosity per star formation rate from Figure \ref{fig:frad_joint_all}. In addition, Figure \ref{fig:fx_constraints} shows the 95 percentile confidence interval and the weighted average\footnote{Here the weighted average is the sum of sample values multiplied by their corresponding weights and normalised to the summed weights.}. The combined constraints reveal a distinct peak in the 1D marginal posterior PDF on the $\log$ X-ray luminosity per star formation rate at $\approx 41.48\ \left(f_X \approx 10\right)$ which is slightly larger than the weighted average at $\approx 40.77\ (f_X \approx 2)$. The 68 percentile confidence interval around the weighted average is 
    \begin{align*}
    &\log_{10} \left(L_X/\mathrm{SFR}\ \mathrm{\left[erg\ s^{-1}M_{\odot}^{-1}yr\right]}\right) \approx 40.77_{-0.90}^{+2.21}\\ 
    &(f_X \approx 2^{+316}_{-1.7}) 
    \end{align*}
    suggesting that early galaxies were between 0.3 and 318 times as X-ray efficient as present-day starbursts. The posterior peak arises from the fact that the 21-cm data disfavours models with low values of $f_X$, while the X-ray Background observations disfavour models with high $f_X$. This example clearly showcases the advantage of using complementary multi-wavelength data in synergy to effectively inform astrophysical models. 
    
    \item $L_\mathrm{r}/\mathrm{SFR}$: The strongest 1D marginal 68 percentile upper limit is seen on the $\log$ radio luminosity per star formation rate, 
    \begin{equation*}
    \log_{10} \left(L_\mathrm{r}/\mathrm{SFR}\ \mathrm{\left[W\ Hz^{-1}M_{\odot}^{-1}yr\right]}\right) \leq 23.32\qquad (f_r \leq 21), 
    \end{equation*}
    suggesting early galaxies were less than 21 times as radio efficient as present-day galaxies. Here the combination of the upper limits from HERA, SARAS 3, and the Radio Background data causes the 1D marginal posterior PDF to go to zero and rule out the highest radio efficiencies at the end of our prior range. This limit is not directly comparable to our previous result \citep[$f_r \leq 32$,][]{bevins_joint_paper} due to the model differences \citep[inclusion of the line of sight fluctuations in the radio background in this work,][]{los_radio_fluc}. 
\end{itemize}

Next, we consider the 2D marginal PDFs of Figure \ref{fig:frad_joint_all} which provide valuable insight into degeneracies of the constrained parameter space.
\begin{itemize}
 \item $L_X/\mathrm{SFR}$ with $L_\mathrm{r}/\mathrm{SFR}$: The clearest constraint is observed in the $L_X - L_\mathrm{r}$ plane. High radio efficiency in combination with low X-ray efficiency result in the strongest 21-cm signals that are the easiest ones to rule out with both HERA and SARAS 3. In addition the Radio Background data rules out the highest radio efficiencies, and the X-ray Background data disfavours the highest X-ray efficiencies at 95\%. We can approximate the 95\% region by the following inequality
    \begin{align}
    \begin{split}
         \log_{10} \left(L_\mathrm{r}/\mathrm{SFR}\right) &\gtrsim -0.04279
         \log_{10} \left(L_X/\mathrm{SFR}\right)^3 \\
        &+ 5.1615 \log_{10} \left(L_X/\mathrm{SFR}\right)^2 \\
        &-206.95 \log_{10} \left(L_X/\mathrm{SFR}\right) + 2782.75.
    \end{split}
    \end{align}  
    \item $f_\mathrm{\ast,II}$ or $f_\mathrm{\ast,III}$ with $L_X/\mathrm{SFR}$ or $L_\mathrm{r}/\mathrm{SFR}$: We see regions of the parameter space corresponding to all the pairs with high Pop II and Pop III star formation together with high radio and X-ray efficiencies being ruled out by more than 95\%. We can loosely define the ruled-out regions ($\gtrsim 95\%$) by the following relations:
    \begin{align}
        \log_{10} \left(L_X/\mathrm{SFR}\right) &\gtrsim -1.1 \log_{10}\left(f_\mathrm{\ast,II}\right)+41.2,\\ 
        \log_{10} \left(L_X/\mathrm{SFR}\right) &\gtrsim -0.8 \log_{10}\left(f_\mathrm{\ast,III}\right)+42.2,\\ 
        \log_{10} \left(L_\mathrm{r}/\mathrm{SFR}\right) &\gtrsim -0.8 \log_{10}\left(f_\mathrm{\ast,II}\right)+23.3,\\ 
        \log_{10} \left(L_\mathrm{r}/\mathrm{SFR}\right) &\gtrsim -0.3 \log_{10}\left(f_\mathrm{\ast,III}\right)+24.2.
    \end{align}
    Such combinations produce strong X-ray or Radio Backgrounds that can be disfavoured by the corresponding observational limits. Additionally, we see suppression in the 2D PDFs in the regions with low X-ray efficiencies and mid to high Pop II and Pop III star formation efficiencies as (combined with high radio efficiencies) these models are disfavoured by the 21-cm data.

\end{itemize}

\subsection{Updated parameter constraints from SARAS 3}\label{sec:updated_saras_constraints}
Naturally, the results presented here are model-dependent. Similar parameter inferences (although excluding X-ray and radio constraints) have previously been conducted using an earlier version of \textsc{21cmSPACE} \citep[][]{bevins_saras3, bevins_joint_paper}. 

Owing to the difference in modelling, we find somewhat different numerical values for the astrophysical parameter constraints with SARAS~3. 
Arguably the most important difference arises due to the star formation prescription with the newly added Pop III star formation and consistent Pop III - Pop II transition \citep{popIII_implement, thomas_popIIIimf}. This important development enabled us, for the first time, to use the SARAS~3 data to constrain Pop III star formation efficiencies. Additionally, the updated models account for the line-of-sight radio fluctuations which cause up to 5\% difference in the global signal compared to the earlier versions of \textsc{21cmSPACE} \citep{los_radio_fluc} softening the constraint on $f_r$.

\begin{figure}
\includegraphics[width=\columnwidth]{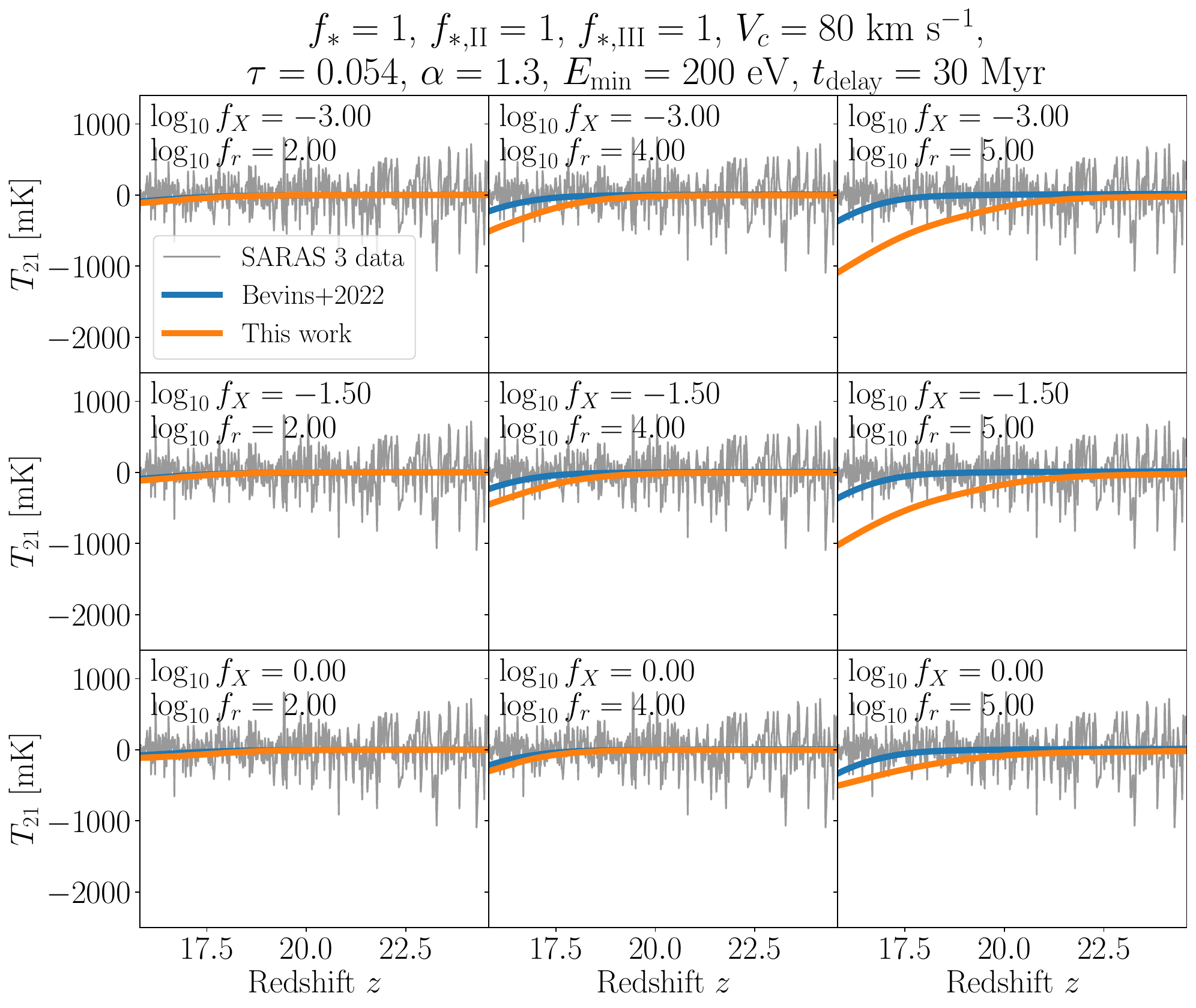}
\caption{Comparison of \texttt{21cmSPACE} global signals within the SARAS 3 band (\mbox{$z_\textrm{SARAS 3}\approx15-25$}) for a previous iteration of \textsc{21cmSPACE} (blue) and the updated version used for this study (orange). For reference we also show the best-fit foreground subtracted SARAS~3 residuals (transparent solid black lines). 
While the updated models include separate star formation efficiencies ($f_\mathrm{*,II}$ and $f_\mathrm{*,III}$), the previous models only had a single star formation efficiency ($f_\mathrm{*}$) and hence only stars with Pop II properties throughout the simulations \citep[e.g.][]{fialkov2014rich, fialkov2018constraining, BevinsSARAS2_2022}. 
Shown are cases with high star formation efficiency and high minimum mass for star formation, for different values of X-ray and radio efficiencies. We find that for low X-ray efficiencies and (relatively) low radio efficiencies, the signals agree and are below the SARAS 3 residuals. However, as we raise $f_r$ the absorption troughs deepen and the signals can be ruled out by the SARAS 3 data with the strongest rejection of models at the lowest $f_X$ values.  
Consequently, we see that with the new models SARAS~3 disfavours low values of $f_X$, a trend that has not been observed in a similar investigation that used the previous version of \textsc{21cmSPACE} \citep{bevins_saras3}.
\label{fig:new_saras_constraint}}
\end{figure}

In our previous studies, the SARAS 3 data did not constrain values of X-ray efficiency \citep{bevins_saras3, bevins_joint_paper}, indicating that the SARAS 3 constraint on $L_X / \mathrm{SFR}$ presented in this work (Figure \ref{fig:fx_constraints}, green curve) is linked to the updated models. 
To understand the new SARAS 3 constraints on low X-ray efficiencies, we compare the global signals produced by the current and the earlier versions of the code. In Figure~\ref{fig:new_saras_constraint}, we inspect high star formation efficiency models for different values of X-ray and radio efficiencies. 
To ensure an as close and fair comparison as possible of the model versions, we look at similar star formation histories by setting \mbox{$f_\mathrm{\ast} = f_\mathrm{\ast,II} = f_\mathrm{\ast,III}=1$}, where $f_\mathrm{\ast}$ is the star formation efficiency parameter in the previous version of the code. 
We find that, with the right timing of star formation (high virial velocity), for low $f_X$ and high $f_\mathrm{r}$ the global signals generated with the previous version of the code (shown in blue) are buried in SARAS~3 residuals (grey), while the global signals generated with the updated version (orange) deepen towards the low-$z$ end of the SARAS~3 band and become brighter than the residuals. The updated models allow for improved modelling of physical processes specific to Pop II and Pop III stars. In the context of the 21-cm global signal this results in more flexibility of the depth and timing of the signal, which leads to constraints from the SARAS~3 data showing a slight disfavouring of low X-ray efficiencies, in contrast to no constraints being seen with the previous version of \textsc{21cmSPACE} models.

\subsection{Functional posterior PDFs of 21-cm observables}
\begin{figure*}
\includegraphics[width=\textwidth]{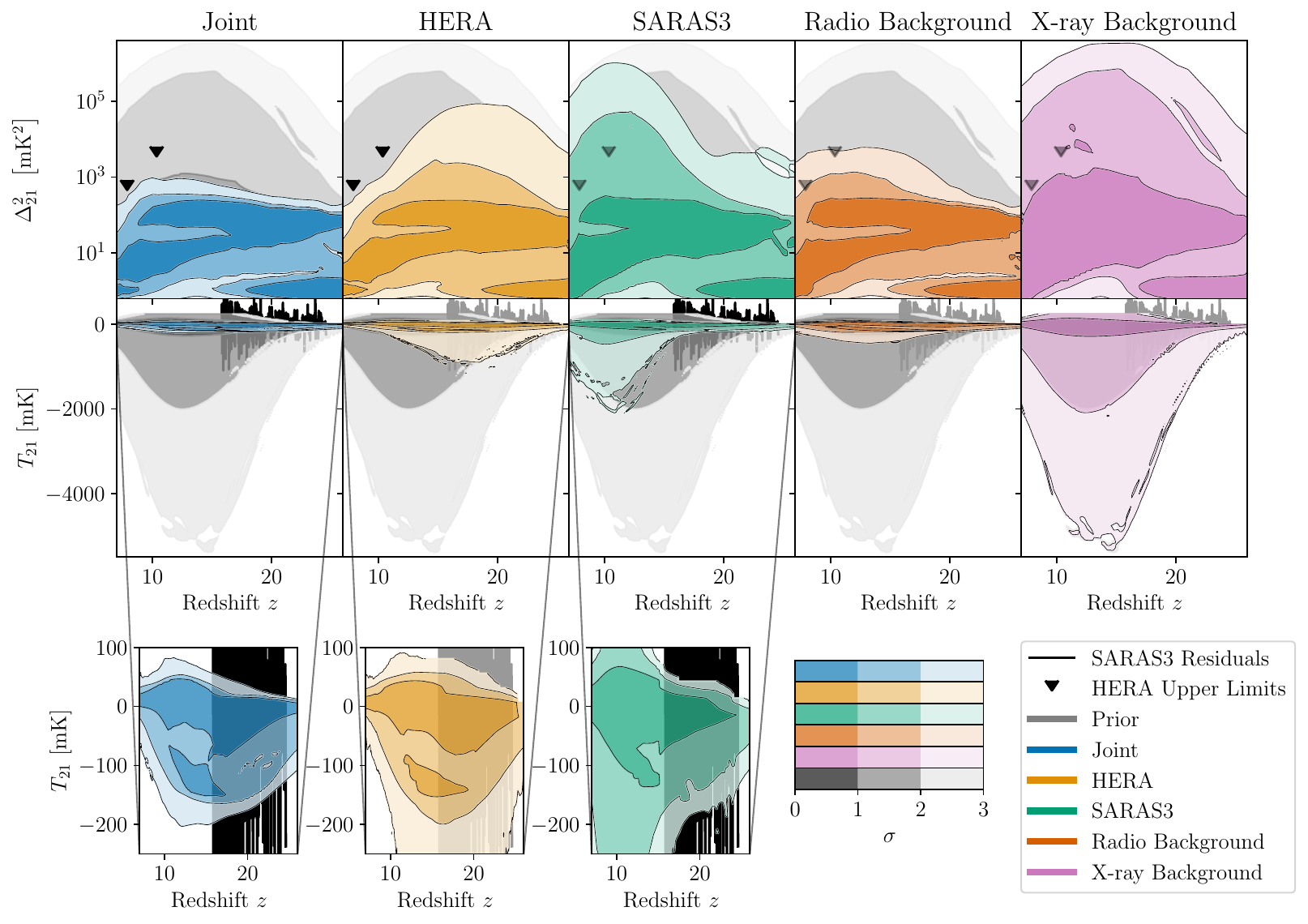}
\caption{Functional prior (grey) and posterior (colored) distributions of the  21-cm power spectrum at $k=0.34~h\mathrm{Mpc^{-1}}$ and the global 21-cm signal at 1 to 3$\sigma$, as indicated by the colorbars. Constraints on the power spectrum space are shown in the top row and global signal space in the middle row; from left to right, we show the joint (blue), HERA (orange), SARAS 3 (green), Radio Background (red) and X-ray Background (pink) constraints. Additionally, zoomed-in plots of the global 21-cm signal posteriors are provided in the third row for clarity for the joint, HERA and SARAS 3 analysis.
For ease of comparison, we also plot the headline HERA 95\% confidence upper limits on the 21-cm power spectrum in each of their redshift bands, $\Delta_{21}^2 \leq 457\ \mathrm{mK^2}$ at $k=0.34~h\mathrm{Mpc^{-1}}$ in the $z_\textrm{Band 2}\approx 7.87$ band, and \mbox{$\Delta_{21}^2 \leq 3496\ \mathrm{mK}^2$} at $k=0.36\ h\mathrm{Mpc}^{-1}$ in the $z_\textrm{Band 1}\approx 10.35$ band~\citep{HERA_IDR3}, these are shown as opaque or transparent when HERA data is included or excluded in the constraint respectively. 
Similarly, the SARAS 3 residuals are plotted in the background as opaque or transparent solid lines when they are included or excluded in the analysis. 
The joint constraints show the most significant compression from the prior to the posterior volume for both types of the 21-cm observables. Models with the power spectra higher than $\Delta_{21}^2 \sim 946\ \mathrm{mK}^2$ at $k=0.34~h\mathrm{Mpc^{-1}}$ and $z\approx10$ are ruled out at $3\sigma$. We also find a posterior global minimum at $z\approx12.24$ suggesting global signals deeper than $\lesssim -203\ \mathrm{mK}$ are ruled out with $3\sigma$ confidence. 
Focusing on the individual experiment constraints, we see that, as expected HERA disfavours models with power spectra above the illustrated upper limits.
Note, the HERA functional posterior 2$\sigma$ contour lies significantly below the shown 95\% confidence upper limits due to three compounding factors: the functional prior favours low powers, the choice of a smoothed-step function likelihood, and the inclusion in our analysis of not just the illustrated limits but also the HERA limits across different observational fields and wavenumbers.
Power spectra at higher redshifts are also constrained by HERA owing to the fact that low and high redshift 21-cm power spectrum magnitudes are correlated. 
Meanwhile, SARAS 3 directly constrains global signals models in $z_\mathrm{SARAS\ 3}\approx 15-25$ observation band, which translates to the limits on the power spectrum in the same redshift range. 
The Radio Background data mainly rules out the highest radio efficiency models, which produce very high amplitude power spectra and deep global signals across the entire redshift range, while the X-ray Background data provides the weakest constraints (posterior is very close to the prior), as it disfavours only high X-ray efficiency models which correspond to low amplitude 21-cm signals.  
\label{fig:dsq_gs_post}}
\end{figure*}

In this section, we examine the prior and posterior functional PDFs in the space of the 21-cm global signals and power spectra shown in Figure \ref{fig:dsq_gs_post}. This allows us to quantify the constraining power of each experiment directly in the space of the observable quantities. It also allows us to illustrate the regions of the prior signal space that are constrained by each type of data. 

\subsubsection*{Constraints on the 21-cm power spectrum}
Top row of Figure \ref{fig:dsq_gs_post} shows the functional prior and posterior PDFs of the theoretical 21-cm power spectra (shown at $k=0.34~h\mathrm{Mpc^{-1}}$) constrained by the combined data set as well as each experiment separately. Among the individual experiments, we see the  prior volume of the power spectrum at $k=0.34~h\mathrm{Mpc^{-1}}$ is strongly contracted by the HERA data. The reduction of the signal space is expected at the redshifts directly observed by HERA ($z_\textrm{Band 2}\approx 7.87$ and $z_\textrm{Band 1}\approx 10.35$). However, owing to the properties of our models (which link high and low-redshift 21-cm signals via structure formation, star formation, heating and ionization histories), HERA limits also contribute to constrain the functional power spectrum posterior at higher redshifts.

As the global signal and the power spectrum are derived from the same underlying astrophysical parameters, constraints on the parameters from one observable will naturally result in constraints on the other observable. Global signals that are ruled out by SARAS3 thus have corresponding power spectra that are also disfavoured owing to the SARAS3 constraints. 
The resulting compression of the power spectrum functional prior in the SARAS 3 band, \mbox{$z_\textrm{SARAS 3}\sim 15 - 25$}, can be seen in Figure \ref{fig:dsq_gs_post}  (green). The strongest impact is in the SARAS~3 band, but, as with HERA, the signals are constrained over a wider redshift range owing to the dependence of the signal on cosmic histories.

The Radio Background data strongly compresses the functional power spectrum prior, ruling out signals across the entire redshift range. Similar to the 21-cm data this is because the Radio Background data directly rule out high radio efficiencies that tend to produce high amplitude power spectra.

Lastly, the X-ray background provides the least constraining power on the functional posterior of the power spectrum PDF. This is because the X-ray limits primarily disfavour high X-ray efficiency models, which have relatively low amplitude 21-cm power spectra. 

It is evident from the figure that the combined data set (blue) achieves the most stringent reduction of the prior space of the 21-cm power spectrum. As we see from the figure, the joint fit indeed combines the constraints from all the experiments. Most notably, the models are constrained across the entire redshift range owing to the complementarity of HERA, SARAS 3, and the Radio Background data. Due to the synergistic approach, no extrapolation of the constraints outside of the observational bands is needed. Using this approach we find the joint $3\sigma$ upper limit on the power spectrum with a maximum (across redshift) of \mbox{$\Delta_{21}^2(k=0.34\ h\mathrm{Mpc^{-1}}) \sim 946\ \mathrm{mK}^2$}  at $z\approx10$. 

\subsubsection*{Constraints on the 21-cm global signal}
We next examine the range of the possible global signals allowed by the data by exploring the functional posterior PDFs depicted in the second row of Figure \ref{fig:dsq_gs_post}. 

The direct limits on the global signal are provided by the SARAS 3 data. In Figure \ref{fig:dsq_gs_post} we show that the SARAS 3 measurements rule out a large portion of the global signal prior space across the whole redshift range explored here, with the most stringent constraints found in the SARAS 3 band ($z=15-25$). Incidentally, the SARAS 3 data provide the strongest available limits on the Cosmic Dawn signals. 

Using the connection between the global signals and the corresponding power spectra, the upper limits on the power spectrum from HERA can be expressed in terms of the constraints on the global signals. The HERA measurements result in the strongest constraints on the low-redshift global signal and, thus, significantly compress the global signal prior volume at the Epoch of Reionization. 

The Radio Background data is also able to rule out a large fraction of the global signal prior space by disfavouring high radio efficiency models that produce deep signals. 

Finally, the X-ray Background data is the least constraining observation in terms of the global signal. This is due to X-ray Background data disfavouring high X-ray efficiency models, which typically correspond to shallow global signals. 

As anticipated, the combined data set provides the strongest constraints, contracting the functional global signal prior PDF significantly to its posterior volume. The details are fully captured in the zoomed-in plot in Figure \ref{fig:dsq_gs_post} (bottom left) where we show that the posterior distribution of the sky-averaged signal has a global minimum at $z\approx12.24$ with signals deeper than $\lesssim -203\ \mathrm{mK}$ being disfavoured at $3\sigma$ confidence.  
\section{Discussion} \label{sec:discussion}
We demonstrated the complementary nature of probes of the early Universe through a joint analysis of data from a 21-cm interferometer HERA, radiometer SARAS 3, as well as limits on excess radio background and the unresolved X-ray background, which yields better constraints on the early astrophysical processes than what can be achieved with each individual experiment. This approach is beneficial already at this early phase of high-redshift observations despite the weakness of the existing limits. 

The importance of the joint analysis will only grow as the quality of data improves. In this study we have used data from the HERA Phase 1 system, however for HERA Phase 2, the antennae correlator and signal chain have been upgraded, and the cross-dipole feeds have been replaced with the new Vivaldi feeds extending the array frequency range to $50-250$ MHz \citep{HERA_PhaseII}. Soon, the upgraded interferometer will have the potential to either detect the signal or lower the upper limits on the 21-cm power spectrum significantly, and it will provide observations at a wider range of redshifts, which will strengthen the constraints on astrophysical model parameters. Likewise, global signal experiments like REACH \citep{reach}, which is already taking data, 
may provide a measurement of the global 21-cm signal with a $\sim 25$ mK RMS noise level \citep{reach}, which is approximately an order of magnitude lower than the calibrated \mbox{$213\ \mathrm{mK}$} RMS value for SARAS~3 \citep{saras3_measurement}. Further, new X-ray missions such as \textit{ATHENA} \citep{athena_xray} and \textit{AXIS}\footnote{https://blog.umd.edu/axis/} and radio observatories \citep[e.g. SKA1-LOW,][]{ska_paper} will provide better measurements of the diffuse backgrounds and, in the latter case, also a potential detection/verification of the 21-cm power spectrum. For completeness, measurements of the neutral fraction (reionization history) and UV luminosity functions \citep[][among others]{park2019_lf} could also be included in a future joint analysis.

Finally, it is important to be aware that the derived constraints are naturally model-dependent. Specifically, the implementation of a separate prescription for Pop II and Pop III stars used in this paper for the first time, allowed us to put some of the first constraints on the properties of Pop III star-hosting galaxies. This is illustrated in section~\ref{sec:results_joint} where the 1D marginal posterior PDF reveals a disfavouring of high $f_\mathrm{\ast,III}$, and the 2D marginal posterior parameter space shows that regions with high $f_\mathrm{\ast,III}$ and high X-ray and radio efficiencies are ruled out. Additionally, the lower limit on the X-ray efficiency from SARAS~3 data was not seen in studies that used a previous version of \textsc{21cmSPACE} \citep{bevins_saras3,bevins_joint_paper}. Compared to previous versions, the updated 21-cm models used in this work can produce deeper and earlier signals for the same low X-ray and high radio efficiencies, and as a result SARAS 3 can disfavour these signals (section~\ref{sec:updated_saras_constraints}). 

Differences also appear when comparing HERA sourced X-ray constraints from \textsc{21cmSPACE} with those of \citet{HERA_IDR3} (Sec. 7.4, Figure 28) using \textsc{21cmFAST} \citep{Mesinger_2011,Murray2020}. The \textsc{21cmFAST} version used in \citet{HERA_IDR3} relies on a star formation prescription extrapolating the suppression of star formation to lower masses and higher redshifts than covered by observations \citep{Tacchella_2013, Mason_2015, Mirocha2016, park2019_lf, Sabti2022}. This suppression delays Cosmic Dawn and the Epoch of Heating relative to \textsc{21cmSPACE} simulations which include small halos that contribute to the X-ray and radio budget earlier. The delay in star formation causes the IGM to finish reionization later, which leads to strong fluctuations and high power within the HERA bands resulting in tighter X-ray constraints. \citet{PopIII_weakens_Xray_bounds} also show that the inclusion of a Pop III hosting galaxy population removes the strong disfavouring of low X-ray efficiencies  \citep[seen in Sec 7.4 of][]{HERA_IDR3}. This is accounted for in the version of \textsc{21cmSPACE} used in this study as we include a separate star formation prescription for Pop II and Pop III galaxies. It is also important to note that the X-ray constraints are reported in two different energy bands e.g. from $E_\mathrm{min}$ to $95$~keV in our case and $<2$~keV in the aforementioned Sec.~7.4 of \citet{HERA_IDR3}.

\section{Conclusions} \label{sec:conclusion}
In this work, we use multi-wavelength data to constrain astrophysical processes at Cosmic Dawn and the Epoch of Reionization by looking at limits on the 21-cm signal, as well as radio and X-ray backgrounds. We reveal a (68\%) disfavouring of Population III star formation efficiencies $\gtrsim5.7\%$. To our knowledge, these are the first constraints of their kind on the first population of stars.  Our other findings are summarised below. 

Similar to our previous work \citep{bevins_joint_paper}, we use the average sky temperature from SARAS 3 \citep{saras3_measurement}. Improving over \citet{bevins_joint_paper}, we use the latest and currently best publicly available upper limits on the 21-cm power spectrum from the HERA collaboration \citep[second public data release from the H1C observing season,][]{HERA_IDR3}. Additionally, we include measurements of the present-day radio background temperature \citep{tradio_data}, and observations of the unresolved integrated X-ray background \citep{hickox2006,harrison2016}. These limits are combined in a fully Bayesian way to yield a joint likelihood that is used for parameter inference. We use these combined data to constrain models generated with a 21-cm semi-numerical code, \textsc{21cmSPACE}, now including a separate prescription for Pop II and Pop III stars, a time-delay from the first Pop III supernova until the onset of Pop II star formation \citep{popIII_implement}, and line-of-sight radio fluctuations in the contribution to the radio background created by early galaxies  \citep{los_radio_fluc}. 

We determine the constraining power of each individual experiment as well as the joint data set, by quantifying the compression from the initial prior volume of the astrophysical parameters to the posterior volume consistent with the data. We find that $47.8^{+0.5}_{-0.5}$\% of the prior volume is consistent with the joint data set. The corresponding percentages for the individual experiments are 
$63.5^{+0.3}_{-0.3}$ for the Radio Background data, $75.5^{+0.4}_{-0.5}$ for HERA, $85.9^{+0.1}_{-0.2}$ for SARAS~3, and $89.7^{+0.2}_{-0.3}$ for the X-ray Background data. As expected, the joint data set provides the tightest constraints. 
When considering limits on the astrophysical model parameters, arguably the most interesting result we find is that the data are sensitive to the properties of first star-forming galaxies. The 1D marginal posterior PDF of the Pop III star formation efficiency, derived from the joint data, suggests a 68\% disfavouring of efficiencies $\gtrsim 5.7\%$. The 2D marginal posteriors expose ruled-out regions of high Pop III star formation efficiency in combination with high X-ray efficiency and high radio efficiency. We also find a preference towards low Pop II star formation efficiencies. While each experiment only slightly favours low Pop II star formation efficiencies, the joint data set provides a 68\% upper limit of $f_\mathrm{\ast,II}\sim 3.5\%$.

Focusing on the X-ray and radio luminosities, we find, in agreement with our previous works \citep[e.g.][]{HERA_IDR2a_limits, bevins_joint_paper}, that 
the 21-cm data disfavour combinations with high radio and low X-ray efficiencies which produce deep global signals and high amplitude power spectra. The novelty of this work is that here we add the upper limits on cosmic X-ray Background and Radio Background in our likelihood formalism. The limit on X-ray Background rules out models with high X-ray efficiencies so that the combined data yield a distinct peak in the 1D marginal posterior PDF of the X-ray luminosity per star formation rate at~\mbox{$\approx 10^{41.48}$ \,erg\,s$^{-1}$\,M$_{\odot}^{-1}$\,yr} and a 68 percentile confidence interval around the weighted mean $\approx 10^{39.87}-10^{42.98}$ \,erg\,s$^{-1}$\,M$_{\odot}^{-1}$\,yr. 
This result suggests that the early galaxies were likely 0.3 to 318 times as X-ray efficient as present-day starbursts. 
Considering radio brightness of early galaxies, we find that the joint fit yields an upper limit on the $\log$ radio luminosity per star formation rate of $\sim 23.32$ (at~68\%). This constraint appears as a result of the joint HERA, SARAS 3, and Radio Background data synergistically compressing the prior space more than any of them individually, and it indicates that early galaxies were less than 21 times as radio-efficient as present-day galaxies. 

Finally, we explore which parts of the initially assumed theoretical signal space are consistent with the data. We present these results in the form of the compression of the functional prior distributions into functional posterior distributions. For the combined data set we find that for the global signal, the functional posterior reaches a global minimum of $\lesssim -203\ \mathrm{mK}$ at \mbox{$z\approx15.24$} ruling out global signals deeper than this limit at $>3\sigma$. Similarly, the functional posterior of the 21-cm power spectrum at $k=0.34\ h\mathrm{Mpc^{-1}}$ reaches a global maximum of $\Delta_{21}^2 \lesssim 946\ \mathrm{mK}^2$ at $z\approx10$ suggesting that signals with a stronger power spectra (at any redshift) are ruled out at $>3\sigma$.

In this work, we have showcased the benefit of synergistic analysis by combining the 21-cm observations with X-ray and radio data. As 21-cm experiments improve and deeper observations and larger surveys from next-generation telescopes become available, the methodology presented in this work will be applied to further the constraints on astrophysical processes at Cosmic Dawn and the Epoch of Reionization. This will allow us to pin down the properties of the first sources of light, build a coherent picture of the infant Universe, and understand how it evolved to its current state.


\section*{Acknowledgements}
SP would like to thank the Cambridge Trust and the Centre for Doctoral Training in Data Intensive Science for their support. 
TGJ acknowledges the support of the Science and Technology Facilities Council (UK) through grant number ST/V506606/1. 
HTJB is supported by the Science and Technology Facilities Council (UK) through grant number ST/T505997/1 and a Kavli Junior Fellowship. 
SH acknowledges the support of STFC (grant code ST/T505985/1), via the award of a DTP Ph.D. studentship, and the Institute of Astronomy for a maintenance award. 
IAC acknowledges support from the Institute of Astronomy Summer Internship Program at the University of Cambridge. 
AF is grateful for the support from the Royal Society through a University Research Fellowship. 
EdLA is supported by the Science and Technology Facilities Council (UK) through his Rutherford Fellowship. 
Sikder and RB acknowledge the Israel Science Foundation (grant No. 2359/20).

The authors would also like to thank Jiten Dhandha for the helpful discussions and reviewing of the pipeline code.

This work was performed using resources provided by the Cambridge Service for Data Driven Discovery (CSD3) operated by the University of Cambridge Research Computing Service (\href{www.csd3.cam.ac.uk}{www.csd3.cam.ac.uk}), provided by Dell EMC and Intel using Tier-2 funding from the Engineering and Physical Sciences Research Council (capital grant EP/T022159/1), and DiRAC funding from the Science and Technology Facilities Council (\href{www.dirac.ac.uk}{www.dirac.ac.uk}).

\section*{Data Availability}

The data generated from the nested sampling is available upon reasonable request to SP. The SARAS 3 data was provided by SS and is available upon reasonable request to SS. 



\bibliographystyle{mnras}
\bibliography{main} 




\appendix

\section{Astrophysical parameter constraints from individual experiments} \label{sec:appendixA}
\begin{figure*}
  \centering
  \begin{subfigure}{0.48\textwidth}
  \centering
    \includegraphics[width=\linewidth]{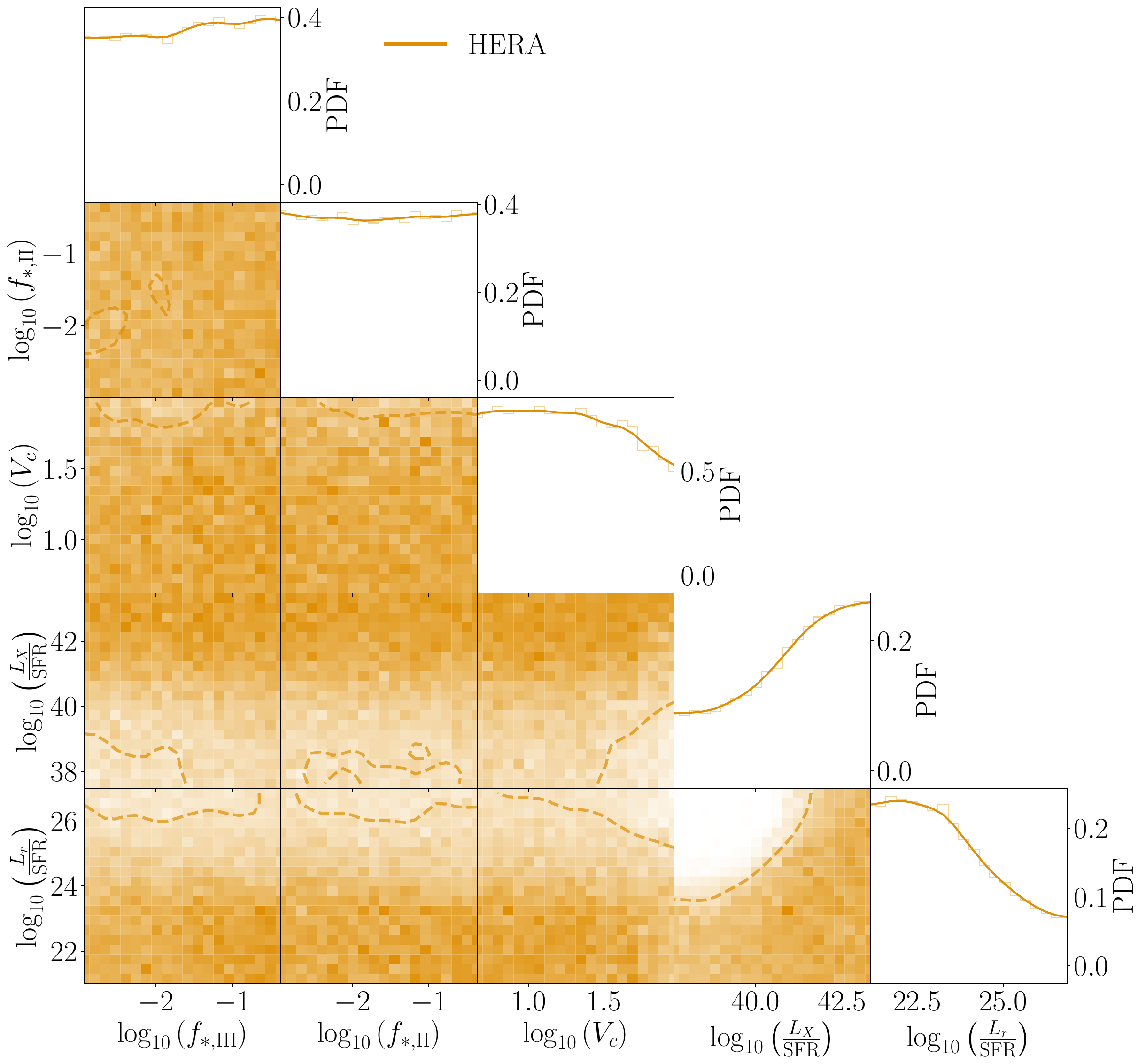}
    \caption{}
    \label{fig:hera_triangle}
  \end{subfigure}
  \begin{subfigure}{0.48\textwidth}
  \centering
    \includegraphics[width=\linewidth]{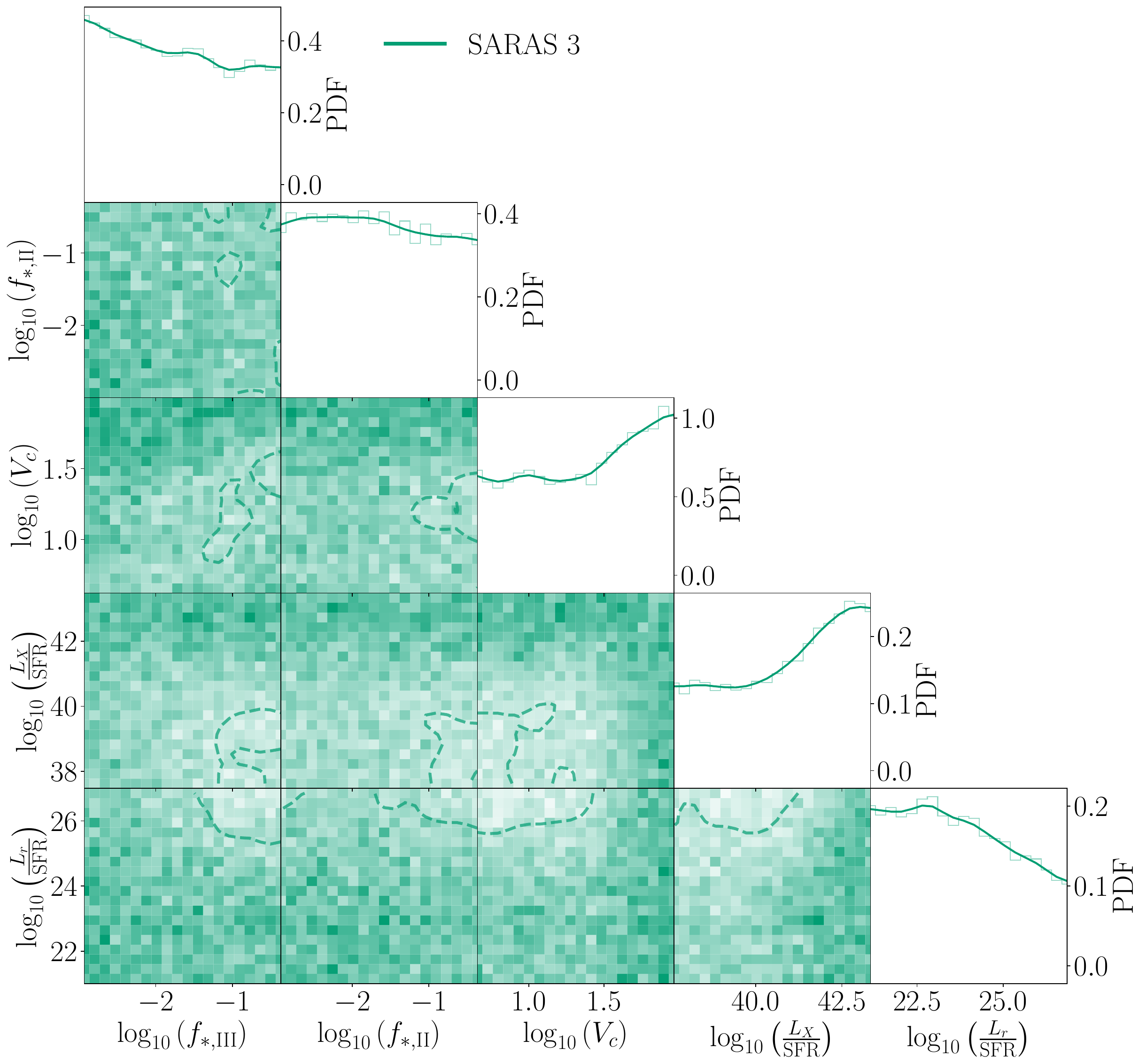}
    \caption{}
    \label{fig:saras_triangle}
  \end{subfigure}


  \begin{subfigure}{0.48\textwidth}
  \centering
    \includegraphics[width=\linewidth]{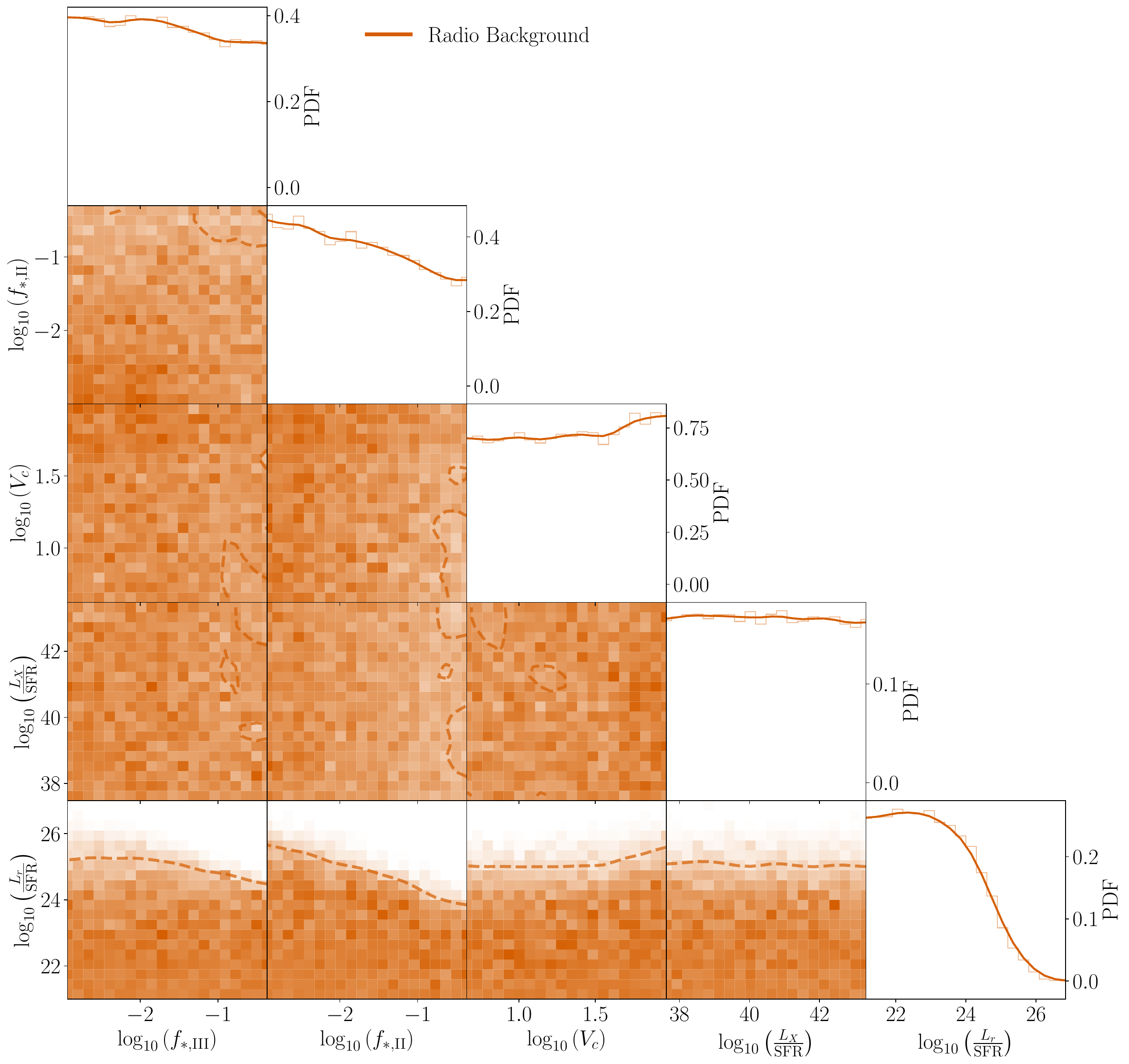}
    \caption{}
    \label{fig:radiobackground_triangle}
  \end{subfigure}
  \begin{subfigure}{0.48\textwidth}
  \centering
    \includegraphics[width=\linewidth]{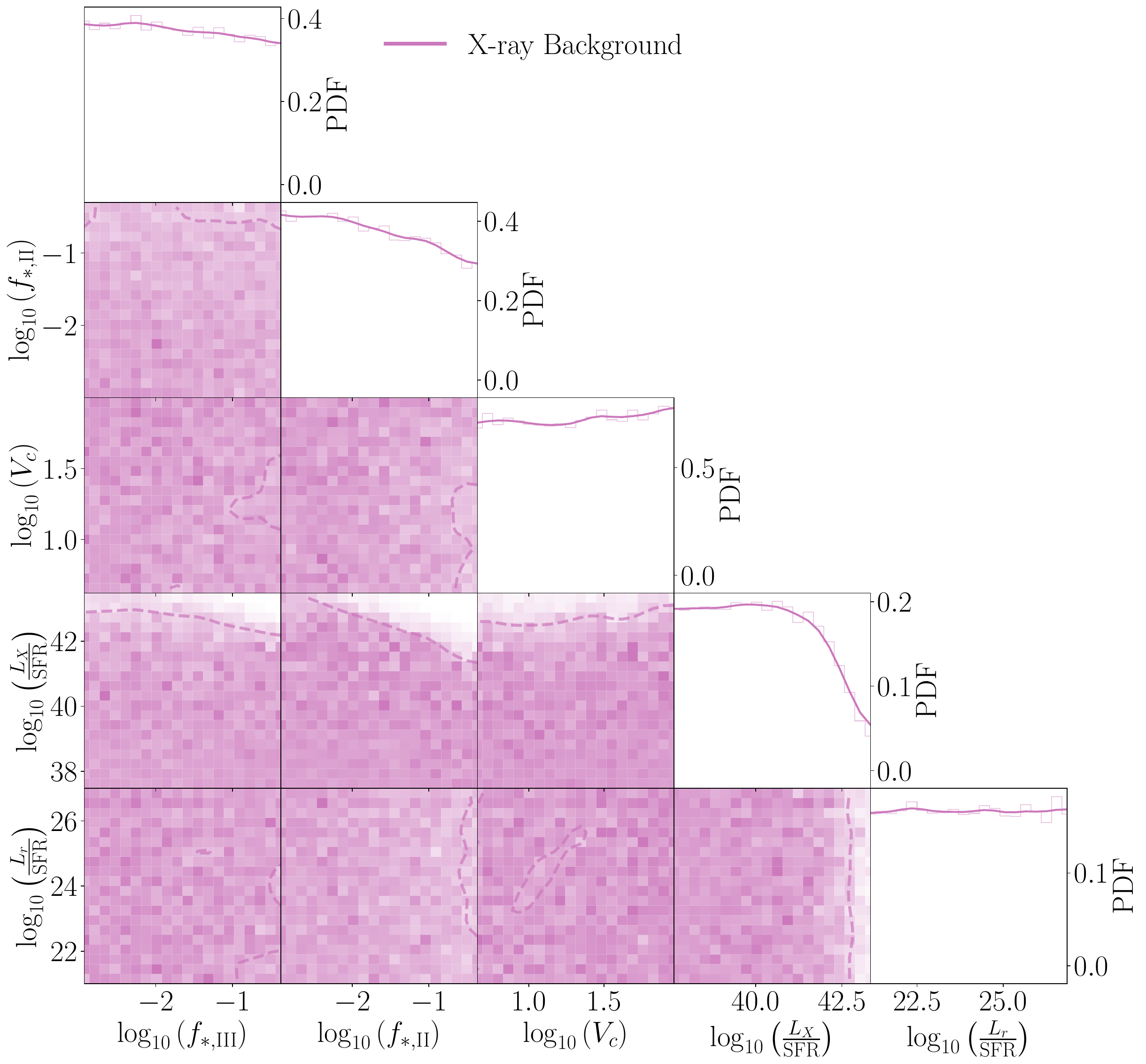}
    \caption{}
    \label{fig:xraybackground_triangle}    
  \end{subfigure}
  \caption{
Marginal astrophysical parameter posteriors inferred from each experiment. 
\subref{fig:hera_triangle} The HERA data \citep[orange, ][]{HERA_IDR3} show 1D marginal posterior PDF limits with a 68 percentile lower limit on the logarithm of the X-ray luminosity per star formation rate, $\geq 40.49\ (38.06)$, and an upper limit on the logarithm of the radio luminosity per star formation rate, $\leq 23.99$. This results in a ruled out region in the X-ray and radio parameter space, as illustrated by the 2D marginal posteriors. 
\subref{fig:saras_triangle} The SARAS3 data \citep[green, ][]{saras3_measurement} reveals 1D marginal posterior PDF limits with the strongest 68 percentile upper limit on the star formation efficiency of Pop III stars, $f_\mathrm{*,III} \leq 5.2\%$. The SARAS 3 data also disfavours models with combinations of low X-ray efficiencies and high radio efficiencies, although less so than the HERA data. This leads to slightly disfavoured regions in the 2D marginal posterior PDFs. 
\subref{fig:radiobackground_triangle} The Radio Background \citep[red, ][]{tradio_data} shows a 68 percentile upper limit on the logarithm of the radio luminosity per star formation rate in the 1D marginal posterior PDF, $\leq 23.56$. This is because the radio efficiency parameter is directly related to the radio background temperature. Additionally, the Radio Background data shows a slight 68 percentile favouring of $f_\mathrm{*,II} \leq 4.6\%$ and $f_\mathrm{*,III} \leq 5.8\%$. The 2D marginals show distinct regions ruled out by the 95\% contour in $f_\mathrm{*,II}$ and $f_\mathrm{*,III}$ versus $L_r/\mathrm{SFR}$ space. 
\subref{fig:xraybackground_triangle} The X-ray Background \citep[pink, ][]{hickox2006,harrison2016} shows a 68 percentile upper limit on the X-ray luminosity per star formation rate in the 1D marginal posterior PDF, $\leq 40.98$, but also slightly favours low star formation efficiencies for Pop III stars, $f_\mathrm{*,III}\leq 6.2\%$, and Pop II stars, $f_\mathrm{*,II}\leq 5.2\%$. The 2D marginal shows a distinct ruled out region in the top right of the $f_\mathrm{*,II}$ versus $L_X/\mathrm{SFR}$ space, but also a disfavouring of high $L_X/\mathrm{SFR}$ with mid to high $f_\mathrm{*,III}$. 
\label{fig:frad_joint_individuals}}
\end{figure*}

In our analysis, we conduct nested sampling runs with constraints from each individual experiment, and all of them jointly imposed. The joint fit, depicted in Figure \ref{fig:frad_joint_all}, provides the strongest constraints on the astrophysical parameter space emphasising the advantages of our multi-wavelength approach. Similar triangle plots were produced for each experiment to illustrate the individual constraints as shown on Figure \ref{fig:frad_joint_individuals}.

\bsp	
\label{lastpage}
\end{document}